\documentclass[10pt]{article}
\RequirePackage[english]{babel}
\RequirePackage{blindtext}
\RequirePackage{lipsum}

\RequirePackage{graphicx}
\RequirePackage{float}
\RequirePackage{subfigure}
\RequirePackage{multirow}
\RequirePackage{multicol}
\RequirePackage{makecell}
\RequirePackage{booktabs}
\RequirePackage{amsmath}

\RequirePackage[version=4]{mhchem}
\RequirePackage{caption}
\RequirePackage{multicol}
\RequirePackage{geometry}

\RequirePackage[utf8]{inputenc}
\RequirePackage[T1]{fontenc}
\RequirePackage{palatino}

\RequirePackage{titling}
\RequirePackage{authblk}
\RequirePackage{hyperref}
\RequirePackage{cleveref}
\RequirePackage{censor}
\RequirePackage{marvosym}

\RequirePackage{soul}
\RequirePackage{tikz}
\usetikzlibrary{calc}

\newenvironment{Table}
  {\par\medskip\noindent\minipage{\linewidth}}
  {\endminipage\par\medskip}

\makeatletter

\usepackage[
    backend=bibtex,
    style=phys,
    sorting=none,
    natbib=true,
    maxnames=5,
    minnames=4]{biblatex} 
\newcommand{\ts}[1]{\textsuperscript{#1}}

\title{Synthetic model of gamma-ray emission during DT experiments on the SPARC tokamak} 

\author{E.~Panontin\ts{1,a}, 
    R.A.~Tinguely\ts{1}, 
    J.L.~Ball\ts{1},
    A.~Grieve\ts{2},
    S.~Mackie\ts{1},
    L.~Nichols\ts{1},
    P.~Raj\ts{2},
    A.A.~Saltos\ts{2},
    L. Singh\ts{1},
    D.~Vezinet\ts{2},
    X.~Wang\ts{1},
    J.C.~Wright\ts{1},
    J.~Rice\ts{1}\\
    \ts{1}Plasma Science and Fusion Center, MIT, Cambridge, MA, USA\\
    \ts{2}Commonwealth Fusion Systems, Devens, MA, USA\\
    \ts{a}Corresponding author: panontin@psfc.mit.edu}

\newcommand{\pfus}{P_{\mathrm{fus}} }
\newcommand{\labr}{\ce{LaBr_3} }
\newcommand{\gammaray}{{$\gamma$-ray} }
\newcommand{\gammarays}{{$\gamma$-rays} }
\newcommand{\subfigsize}{0.45}

\begin{document} 
\maketitle

\section*{Abstract} 
In thermonuclear plasmas, plasma ions undergoing nuclear reactions emit gamma-rays with energies in the MeV range. Their spectroscopy can convey much plasma information, such as the DT fusion power, the spatial and velocity distributions of the fast ions, and the plasma heating performance. In the present work, we simulate the gamma-ray emission expected in the SPARC tokamak during a primary reference discharge, when the tokamak is expected to generate $140$ MW of fusion power and reach an energy gain factor of $Q\approx11$. We focus particularly \ce{T(D, $\gamma$)^5He}, \ce{^{10}B(^{4}He, p $\gamma$)^{13}C} and \ce{D(^{3}He, $\gamma$)^{5}Li} reactions. We use realistic plasma profiles calculated with the TRANSP code and simulate radiofrequency heating of the plasma with CQL3D and TORIC. Possible locations for gamma spectrometers based on lanthanum bromide inorganic scintillators are suggested. For each, the signal-to-noise ratio of gamma-rays over neutrons is evaluated using the ray-tracing code ToFu and high fidelity Monte Carlo models (MCNP and OpenMC) to solve radiation transport in SPARC. A dedicated neutron attenuator made of high density polyethylene is scoped to allow gamma-spectroscopy during high neutron yield experiments. And finally, the performance of  \ce{LaBr_3} detectors in reconstructing the fusion power generated by SPARC is discussed.

\begin{multicols}{2}

\section{Introduction}\label{sec:intro}
In thermonuclear plasmas, nuclear reactions within the ion population, composed of both fuel species and impurities, can generate a variety of different \gammarays spanning a wide energy range, $[0.1, 20]$ MeV.
\gammaray detection has been conducted in both magnetic and inertial confinement experiments, proving their importance as high temperature plasma diagnostics.
The most notable application is the reconstruction of the DT fusion power generated on the National Ignition Facility and the Joint European Torus using the measurement of DT \gammaray emission~\cite{meaney2021, dalmolin2024}. 
On JET, a $19$ channel $\gamma$-camera also enabled experimental reconstruction of the fast ions poloidal distribution, benchmarking first principle simulations~\cite{panontin2021, panontinphd, fugazza2026}. 
In low emissivity scenarios, high resolution detectors such as high purity germanium could resolve the Doppler shift of characteristic spectral lines. 
From this, the velocity distribution of fast ions, such as the target of the Ion Cyclotron Resonance Heating (ICRH) power, was reconstructed~\cite{curuia2017, nocente2020}.

These results were made possible by the advent of \labr inorganic scintillation crystals~\cite{cazzaniga2013, cazzaniga2015, nocente2016, rigamonti2018}, which have a high neutron hardness and are capable of measuring rates up to the MCps range with a good energy resolution (as good as $2.8$\% at \ce{^{137}Cs}, $\mathrm{E}_{\gamma} = 662$ keV). These detectors are currently the best candidates for conducting \gammaray spectroscopy in experiments where plasmas will reach break-even conditions, such as SPARC, ARC and ITER~\cite{nocente2017}.

SPARC is a high-field tokamak~\cite{creely2020}, currently under construction in Devens, MA, by Commonwealth Fusion Systems. 
The device has been designed to produce up to $\pfus=140$ MW of fusion power from a DT plasma and is expected to multiply the energy necessary to sustain the plasma up to a factor of $Q\approx11$~\cite{creely2020}. 
If successful, SPARC will pave the way to the ARC tokamak: a $400$ MW fusion power plant fueled with a DT plasma mixture and connected to the grid~\cite{hillesheim2026}.
An ambitious milestone of the SPARC project is to reach $Q>1$ in its first plasma campaign. 
In support of this milestone, SPARC plans to install a comprehensive neutron diagnostic suite~\cite{raj2024, mackie2024, ball2024, dallarosa2024, lobelo2025}. 
The main mission of these systems is to provide a reliable estimate of $\pfus$ through a redundant measurement of the DT neutron yield. 

In this context, it would be valuable to introduce an alternative and independent method to infer $\pfus$ based on a completely different physics measurement. 
Such a diagnostic would reduce the effects of systematic errors in the inference of $\pfus$, thus increasing the overall accuracy of the analysis. 
A strong candidate is, indeed, gamma spectroscopy. 
The structure of the SPARC facility could in principle host a multi-line of sight (LOS) gamma diagnostic similar to the one installed on JET.
The only question is: what kind of \gammaray diagnostics would best support the SPARC mission?

\begin{Table}
\centering
\captionof{table}{\label{tab:gamma:reactions} List of nuclear reactions that emit \gammarays of interest for DT operations on SPARC. The average energy of the gamma emission is also reported.}
\smallskip
\begin{tabular}{l r}
\toprule
Reaction & E$_\gamma$ [MeV]\\
\midrule
\ce{D(D,$\gamma$)^4He} & 23.8\vspace{3pt}\\

T(D,$\gamma$)\ce{^5He} & 16.7\\
 & 13.5\vspace{3pt}\\
\ce{D(^3He,$\gamma$)^5Li} & 16.4\vspace{3pt}\\
\ce{^{10}B}(\ce{^4He},p $\gamma$)\ce{^{13}C} & 3.09\\
 & 3.68\\
 & 3.85\\
\bottomrule
\end{tabular}
\end{Table}

In the present work, we investigate the opportunities, challenges, and limits of measuring \gammaray emission in SPARC~\cite{creely2020} during DT operation. 
We focus on the primary reference discharge (PRD) scenario~\cite{creely2020} ($Q=11$, $B = 12.2$ T, $n_e = 3.1 \times 10^{20}$ \ce{m^{-3}}, $n_{DT}/n_e = 0.85$ with 50-50 mixture of D-T, $I_P= 8.7$ MA, $P_{\mathrm{ICRH}} = 11.1$ MW, and $P_{\mathrm{FUS}} = 140$ MW), during which SPARC will operate at maximum fusion power. We also consider a Q>1 scenario, during which SPARC will attempt to reach $Q\approx1$ and produce $P_{\mathrm{FUS}} \approx 10$ MW.
A list of \gammaray of interest for this plasma scenario is reported in~\cref{tab:gamma:reactions}.
As mentioned, the \ce{T(D, $\gamma$)^{5}He} fusion reaction could be used as a second independent fusion power measurement, similarly to what has been recently done on NIF~\cite{meaney2021} and JET~\cite{dalmolin2024}. 
The \ce{D(^3He,$\gamma$)^5Li} fusion reaction has been previously measured at JET to study fast ions generated by ICRH in \ce{D^3He} experiments~\cite{cecil1985, nocente2020, panontin2021}. 
Similar measurements could be performed on SPARC during DD experiments when ICRH targets the \ce{^3He} minitory.
During PRD experiments, on the other hand, this reaction represents a background contribution to the \ce{T(D, g)^{5}He} \gammaray.
The \ce{D(D,$\gamma$)^4He} reaction has been previously proposed as an independent measurement of the DD fusion power output~\cite{cecil1985:2}.
We also investigate the \ce{^{10}B}($\alpha$, p $\gamma$)\ce{^{13}C} reaction for diagnosing of the slowing down distribution of fusion-born alpha particles for burning plasmas studies and transport code validation~\cite{kiptily2018, gallmann1969, bonner1956}.
In~\cref{sec:emissivity} we discuss the \gammaray yield for each of these reactions. 

\textbf{
\begin{figure}[H] 
\centering
\includegraphics[scale=0.1]{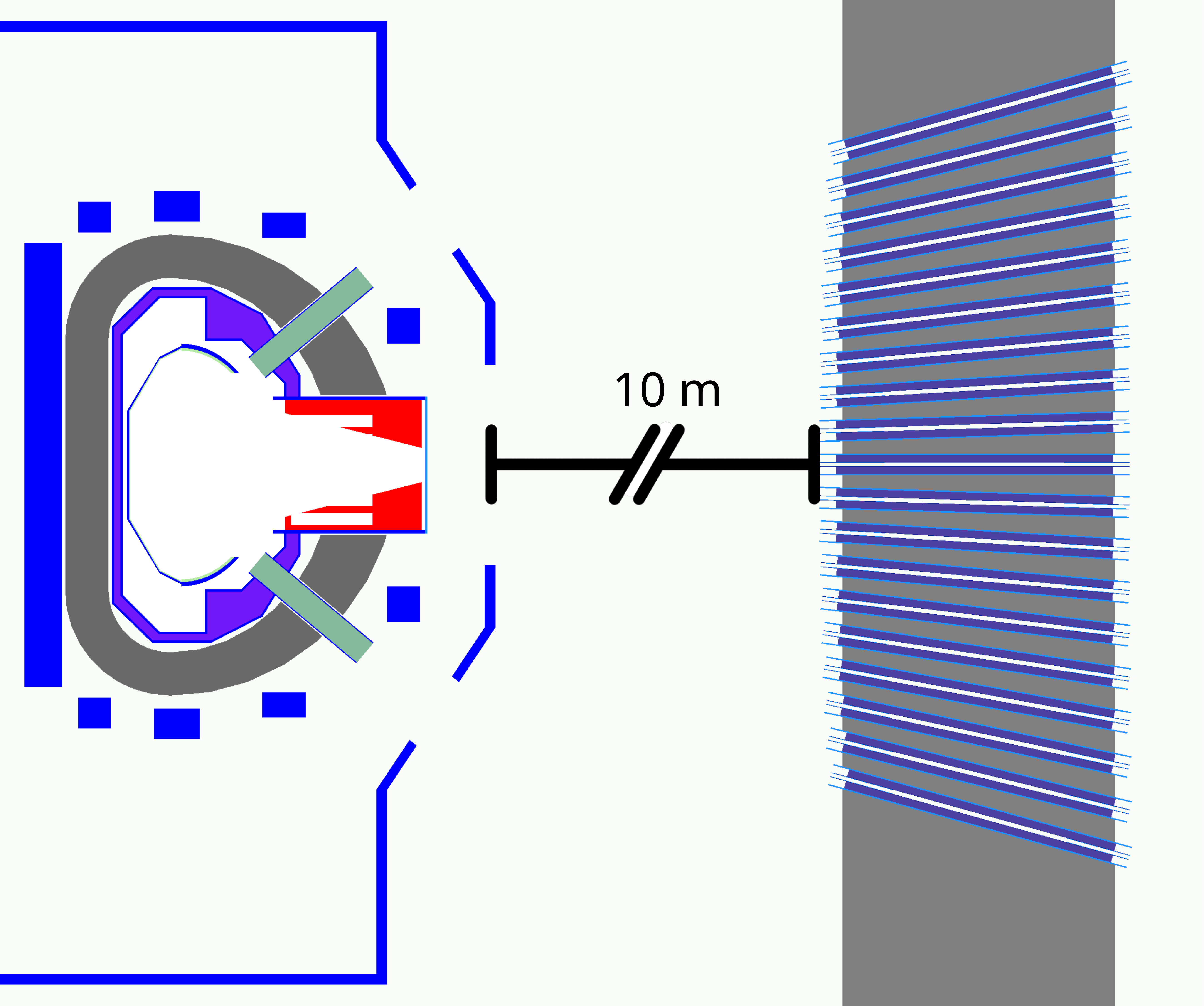}
\caption{Geometry of SPARC and NCAM LOS as implemented in the OpenMC model (not to scale). More information about the full model can be found in refs.~\cite{wang2024, wang2025}.}
\label{fig:SPARC}
\end{figure}}

Starting from its first campaign, SPARC will be equipped with a hard X-ray (HXR) monitor~\cite{vezinet2024, panontin2024}, which will use a \labr crystal to measure photons in the MeV energy range. 
This diagnostic will primarily be responsible to measure \gammarays emitted by {\it bremsstrahlung} interactions between runaway electrons (RE), the rest of the plasma, and SPARC first wall. It will focus on plasma start-up, with the possibility to study also post-disruption RE.
The HXR monitor will also be sensitive to \gammarays emitted in nuclear reactions between plasma ions. 
However, the detector has a wide field of view covering the whole poloidal and toroidal cross-section of the toakamak and can mount only up to 80 cm of neutron attenuation material.
This results in an expected neutron rate in excess of the spectroscopic capabilities of \labr detectors (roughly $5\times10^5$ Cps) during plasma flat-top, making it unsuited to perform \gammaray spectroscopy. 
As an alternative location, we propose to consider the SPARC neutron camera (NCAM), that will observe the plasma behind 19 lines of sight (LOS)~\cite{raj2024, ball2024, mackie2024, dallarosa2024} with a clear field of view (\cref{fig:SPARC}). 
We investigate the possibility of performing \gammaray spectroscopy with traditional \ce{LaBr_3} inorganic scintillators installed on vacant NCAM lines of sights or, alternatively, behind the neutron detectors.
In~\cref{sec:emissivity}, the \gammaray rates in each LOS are evaluated using a deterministic optical ray-tracing code, ToFu~\cite{vezinet2016}, as a simple and fast approximation tool for well-collimated lines of sight. 
A similar study of the DT \gammaray emission on SPAARC has been independently aconducted in ref.~\cite{fugazza2026:1}. 
The present work presents a broader scope of \gammaray signals than ref.~\cite{fugazza2026:1}, as we consider also \ce{D^3He}, $\alpha$\ce{^{10}B} and DD reaction. 
For what concerns DT signals, the two works agree on the \gammaray rates emerging from the collimator system. 

However, the two works differ on the method adopted to estimate the neutron-induced background at the detector. 
In turns, they induces different attenuation strategies and measured signal levels.
In ref.~\cite{fugazza2026:1}, the prompt-gamma ray coming from the torus hall are estimated by rescaling the data measured on JET to the case of SPARC.
While, in~\cref{sec:bkgrd}, we study the neutron background at the proposed detector position using Monte Carlo codes for radiation transport, such as MCNP~\cite{werner2018} and OpenMC~\cite{romano2015}.
In particular we leverage a high fidelity OpenMC model of SPARC~\cite{wang2024, wang2025} to simulate the flux of neutron and prompt-gamma that would reach the proposed detector position during DT operations.

From dedicated neutronics simulations, a neutron attenuator based on high density polyethylene is scoped to enable \gammaray spectroscopy in high neutron backgrounds (\cref{sec:bkgrd:attenuator}). 
We consider a large, $3$ inch $\times 6$ inch (diameter $\times$ height), cylindrical \labr detector placed behind the collimator and its response function to the various radiation is calculated using MCNP (\cref{sec:signal}).
Such a detector size was proven to give superior results, when compared with smaller ($1 \times 2/3$-\ce{inch^2}) detectors~\cite{nocente2020, panontin2021, dalmolin2024, rebai2024, marcer2025}, and thus can best support \gammaray spectroscpy on SPARC.
Finally, the first estimate of a spectrum measured by a \labr detector during SPARC DT operations is presented in \cref{sec:signal}. 
The challenges of background subtraction to isolate the contribution of DT \gammarays are discussed, and the statistics are calculated for different values of $\pfus$. 


\section{\label{sec:emissivity}Gamma-ray and neutron sources}
\gammaray emissivity $Y_\gamma$ [$\gamma$/s/m$^3$] in any point inside the plasma can be calculated {\it via}:
\begin{eqnarray}\label{eq:emissivity} 
Y_\gamma &=& \frac{1}{1+\delta_{ij}} \, n_i \, n_j \nonumber\\
        & &\times\int_{v_i, v_j} f_i(v_i) \, f_j(v_j) \, v_{\mathrm{rel}} \, \sigma_\gamma(v_{\mathrm{rel}}) \, \mathrm{d}\! v_i \, \mathrm{d} \! v_j,\nonumber\\
&=& \frac{1}{1+\delta_{ij}} \, n_i \, n_j \left<v_{\mathrm{rel}} \, \sigma_\gamma\right>_{v_{\mathrm{rel}}}.
\end{eqnarray}
$n_i$ and $n_j$ [1/\ce{m^3}] are the densities of the two ion species involved in the nuclear reaction; the Kronecker delta $\delta_{ij}$ is used to avoid counting the same particle twice when $i=j$, {\it i.e.} if they are the same species. $f_i$ and $f_j$ are the velocity distributions of the two species, they are normalized to $1$ when integrated over all velocity dimensions. $v_{\mathrm{rel}} = \| v_i-v_j\|$ [m/s] is the relative velocity between the two particles and $\sigma_\gamma$ [\ce{m^2}] is the cross-section of the nuclear reaction, expressed as a function of $v_{\mathrm{rel}}$. Finally, the integral $\left<v_{\mathrm{rel}} \, \sigma_\gamma\right>_{v_{\mathrm{rel}}}$ is also called the reactivity.

In the remainder of this section, we solve equation (\ref{eq:emissivity}) to calculate the \gammaray emissivity for different nuclear reactions in a SPARC poloidal section. We use realistic plasma profiles simulated using the TRANSP code~\cite{rodriguez2022} and CQL3D+TORIC~\cite{lin2020}. In particular, ref.~\cite{rodriguez2022} simulates a PRD plasma and predicts a plasma performance of 110 MW fusion power. In order to scan different plasma performance (e.g. the Q>1 scenario), we assume that the total DT neutron as well as the \gammaray yields scale linearly with the fusion power. Using this approach we can rescale the \gammaray emissivity for PRD-like plasmas generating up to the 140 MW of power~\cite{creely2020}. Figures in~\cref{sec:emissivity,,sec:bkgrd,,sec:signal} assume 140 MW of power for the PRD scenario and 10 MW of power for the Q>1 scenario.

The fraction of photons reaching the detector positions at the end of the NCAM collimators is then calculated using the ToFu code~\cite{vezinet2016}. ToFu is a ray-tracing code that uses optical propagation to estimate the photon flux seen by a detector installed on an unobstructed, collimated LOS. For these calculations, we assume the gamma source to be toroidally axisymmetric, and the scattering through the materials that compose the LOS to be negligible. A validation of the neutron flux predicted by ToFu at the NCAM detector position versus full Monte Carlo simulations performed with OpenMC is reported in~\cite{wang2025}. ToFu is shown to agree with OpenMC if the LOS are filled with vacuum; moreover, the neutron spectrum simulated by OpenMC confirms that the scattered component will be negligible.

\begin{figure*}[!t] 
\centering
\subfigure[]{\label{fig:transp:n}
\includegraphics[width=0.45\linewidth]{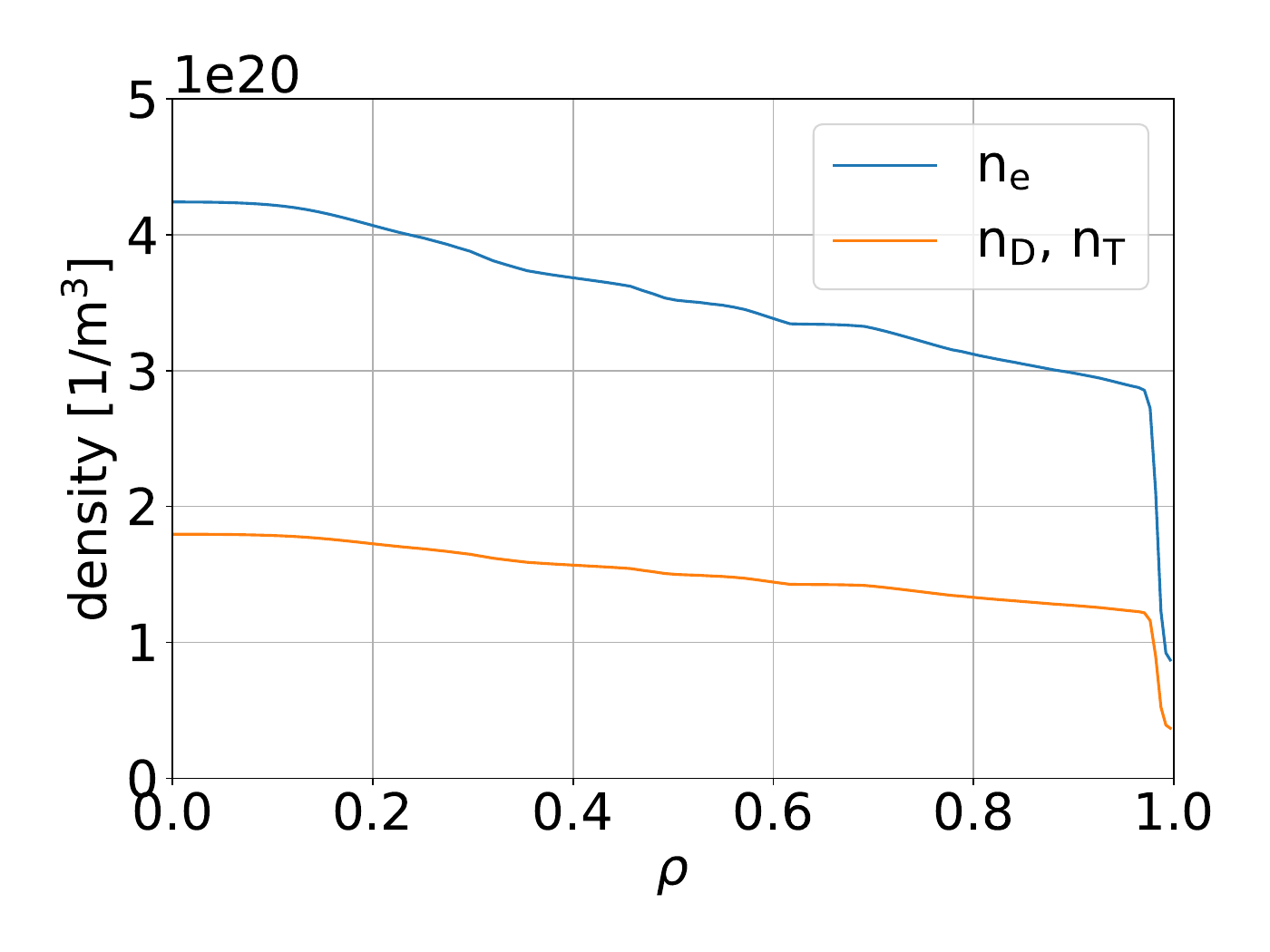}
}\subfigure[]{\label{fig:transp:t}
\includegraphics[width=0.45\linewidth]{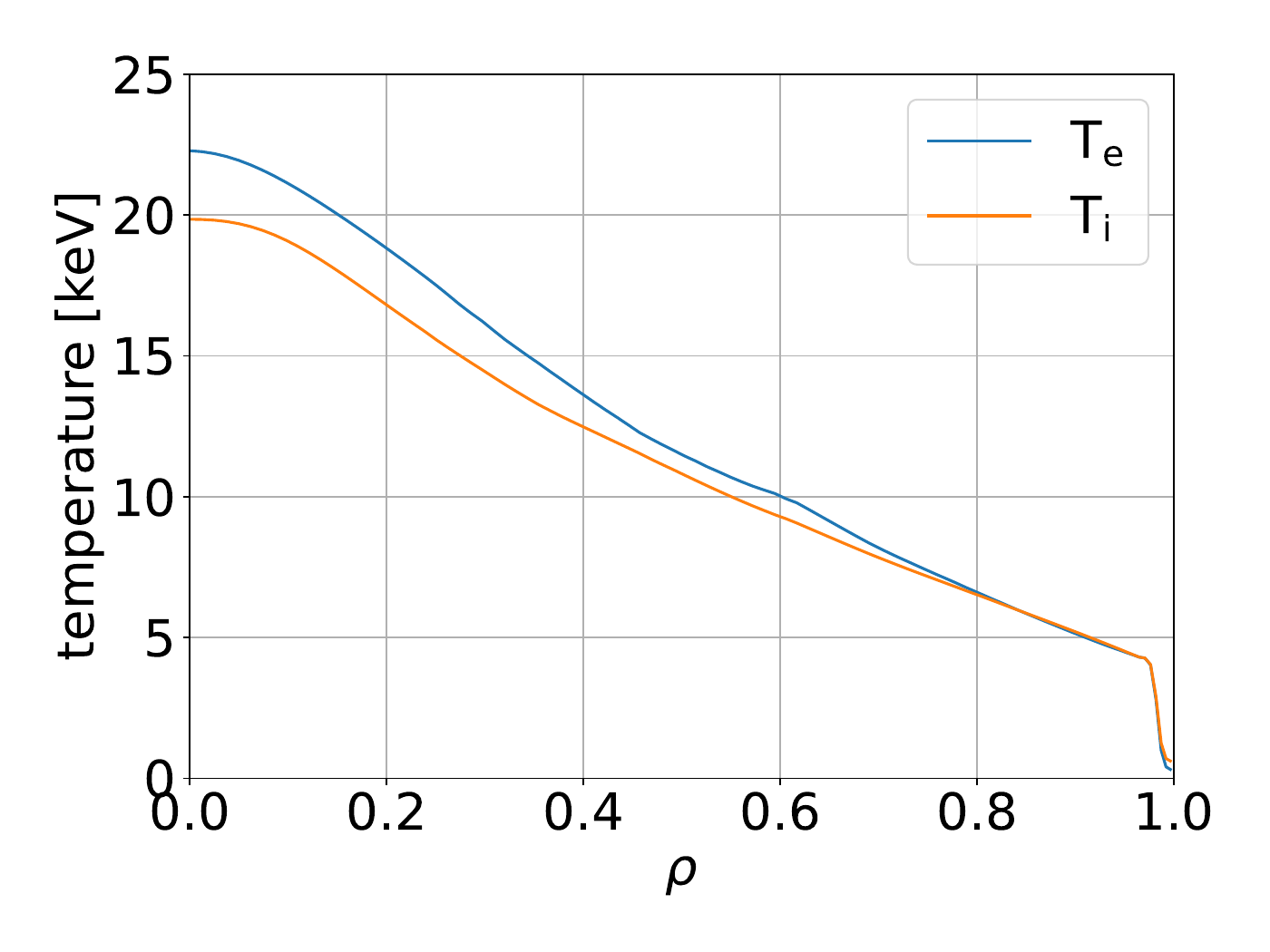}
}\\
\caption{Plasma profiles from ref.~\cite{rodriguez2022} for a SPARC plasma generating 110 MW of fusion power. \subref{fig:transp:n} electron (\ce{n_e}), deterium (\ce{n_D}) and tritium (\ce{n_T}) densities. \subref{fig:transp:t} electron (\ce{T_e}) and ion (\ce{T_i}) temperatures.}
\label{fig:transp}
\end{figure*}

\begin{figure*}[!b] 
\centering
\subfigure[]{\label{fig:DT:emissivity:poloidal}
\includegraphics[width=0.33\linewidth]{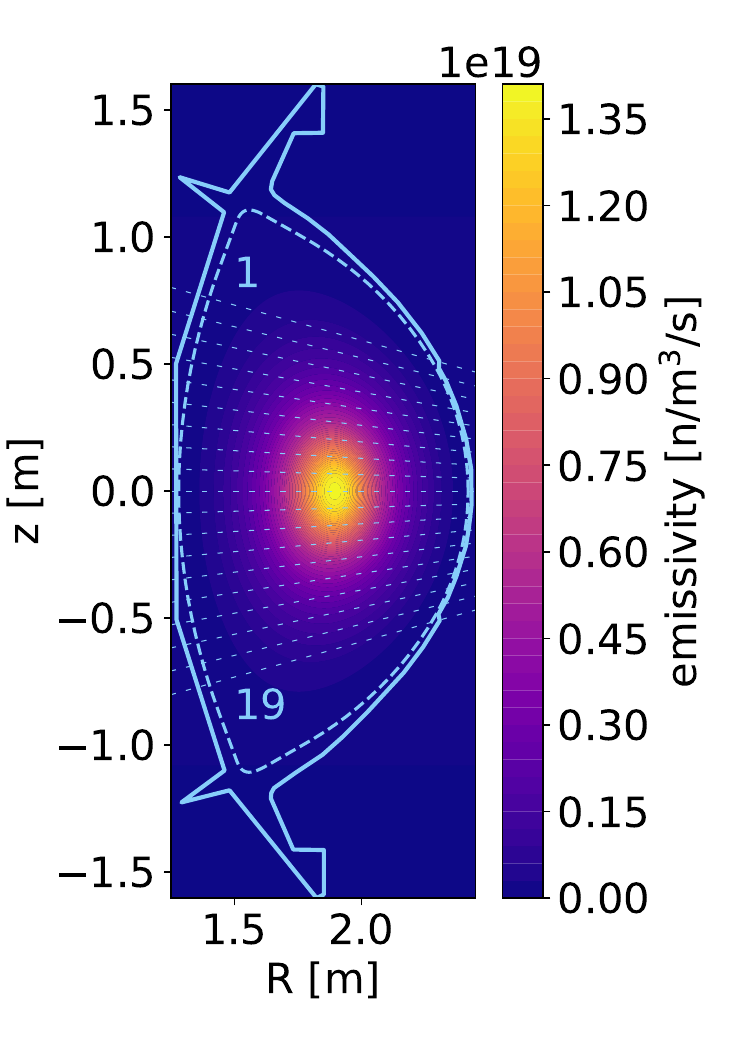}
}\subfigure[]{\label{fig:DT:emissivity:ncam}
\includegraphics[width=0.58\linewidth]{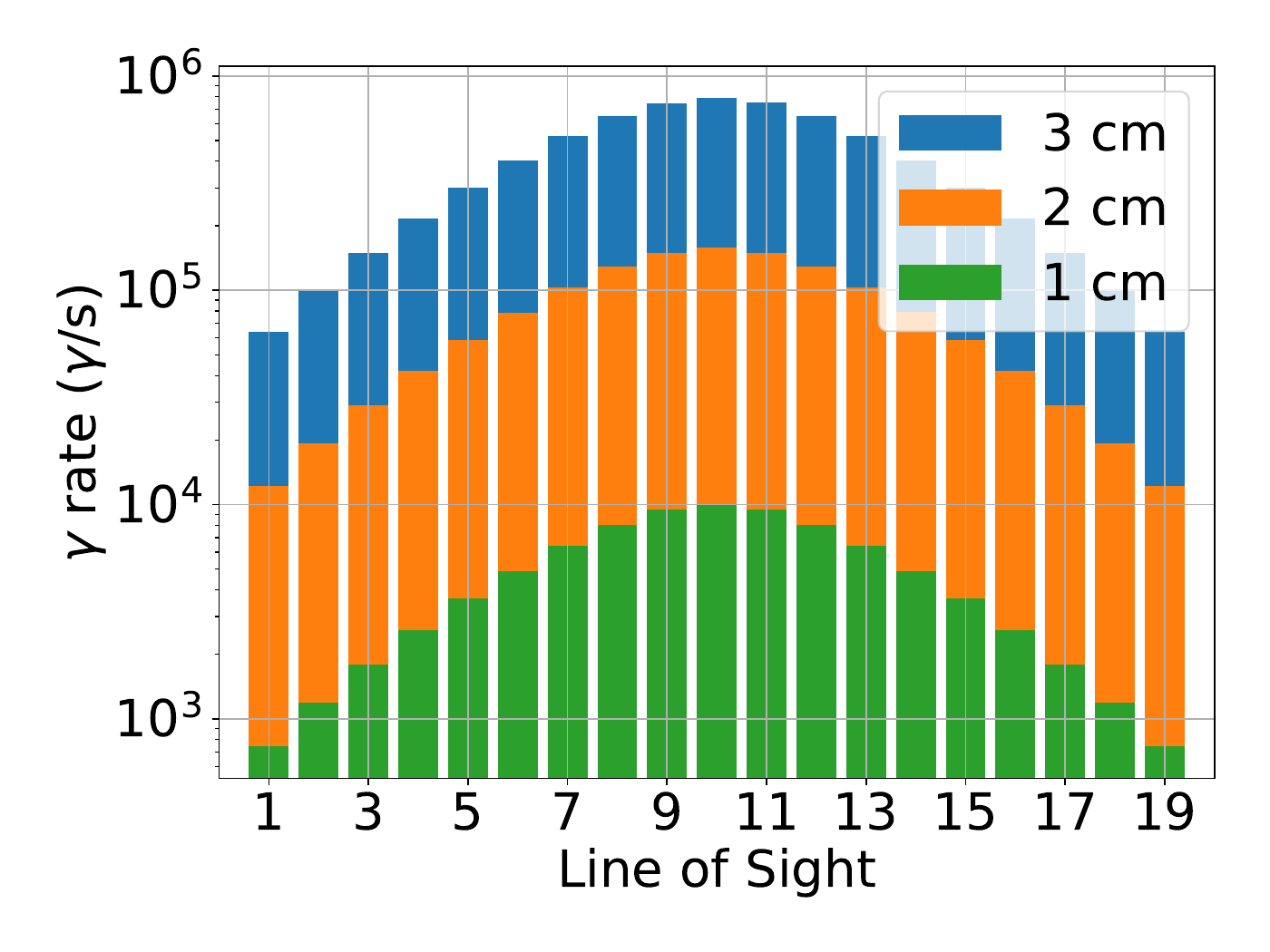}
}\\
\caption{\subref{fig:DT:emissivity:poloidal} Poloidal profile of plasma yield for the D+T fusion reaction (both n and \gammaray branches included). First wall (solid line), LCFS (dashed line) and NCAM LOS (dashed lines) are shown with light-blue with top and bottom LOS labeled. \subref{fig:DT:emissivity:ncam} Expected rate of DT born, 16.7 MeV and 13.5 MeV \gammarays at the end of the NCAM collimators as calculated with ToFu. Three collimator diameters are considered: $1$ cm (green), $2$ cm (orange), $3$ cm (blue).}
\label{fig:DT:emissivity}
\end{figure*}

The geometry of the SPARC tokamak, and NCAM collimators as implemented in OpenMC are shown in \cref{fig:SPARC} for clarity. The NCAM collimators are approximately $2.5$ m long holes casted in the concrete wall between the SPARC torus hall and the diagnostic laboratories. The collimator diameter is defined by custom made inserts and can be changed between campaigns depending on the expected fluxes and the detector geometry deployed in each channel. For \gammaray diagnostics, three different collimator diameters $D = 1$, $2$ and $3$ cm are considered. The fluxes returned by ToFu are obtained assuming a cylindrical detector located at 10 cm from the end of the collimator with its axis parallel to the LOS. The diameter of the detector implemented in ToFu is large enough for the detector to intercept the entire solid angle defined by the collimators and the flux over the wetted area can be considered uniform. Since the wetted area is proportional to the square of the collimator diameter, then ToFu results are here reported in units of total particles reaching the detector per second, to ease the comparison of different collimator diameters. 

\subsection{DT fusion reaction}\label{sec:emissivity:dt}
Most of the current designs of fusion reactors, intend to produce nuclear energy using a DT fuel mix, which is the most reactive fuel at a temperature of about 10 keV. The gold standard to reconstruct the total DT fusion power generated by a machine involves the measurement of the $14.1$ MeV neutrons produced by such reactions~\cite{gatu2022, andersson2009, raj2024, mackie2024, ball2024}. However, the DT fusion reaction can also produce two \gammarays, which could be measured with an absolutely calibrated detector to reconstruct the total \gammaray yield in the plasma following a procedure similar to the one studied in ref.~\cite{mackie2026}. From the \gammaray yield one can reconstruct $\pfus$ knowing the gamma-ray-to-neutron {\it branching ratio} for the DT reaction, i.e. the probability for the reaction to emitt a \gammaray divided by the probability of emitting a neutron. This is known with limited precision: in the present work, we adopt the branching ratio reported in ref.~\cite{dalmolin2024} ($(2.4\pm0.5)\times10^{-5}$), which has an uncertainty of about 20\%. Moreover the measurements of the branching ratio performed by different experiments differ by up to a factor $20$ (see fig. 4 in ref.~\cite{dalmolin2024}).  The advantages of the analysis conducted in ref.~\cite{dalmolin2024} are several: it relies on spectroscopic measurements of the DT \gammaray spectrum; the neutron-induced background was lower than other works that conducted similar spectroscopic measurements; the authors considered both \gammarays emitted in the DT reactions; the measured spectrum was fit using a R-matrix model for the DT \gammaray source spectrum. It is also worth noting that ref.~\cite{dalmolin2024} measured the DT branching ratio on a magnetic confinement fusion device with a detector technology similar to what we consider in~\cref{sec:signal}. 

The complete scheme of the DT fusion reaction, considering both the neutron and the \gammaray branches, can be written as:
\begin{equation}\label{eq:dt}
    \ce{D}+\ce{T}
    \begin{cases}
        \overset{\approx 1}{\longrightarrow} &\alpha(3.5~\mathrm{MeV})+\ce{n}(14.1~\mathrm{MeV}),\\
        \overset{\approx 2.4\times10^{-5}}{\longrightarrow}&
            \begin{cases}
                ^5\mathrm{He}+\gamma_0(16.7~\mathrm{MeV}),\\
                ^5\mathrm{He}+\gamma_1(13.5~\mathrm{MeV}).                
            \end{cases}        
    \end{cases}
\end{equation}
The measurement of the \gammarays emitted by DT fusion reactions is complicated by the fact that the two \gammarays are emitted with a broad energy distribution centered on $E_{\gamma_0} = 16.7$ MeV and $E_{\gamma_1} = 13.5$ MeV. In this work we adopt the spectral shape reconstructed in ref.~\cite{rebai2024,dallarosa2024} using the R-matrix method. According to this model, the relative yield of $\gamma_1$ relative to $\gamma_0$ is about $1.09 \pm 0.25$.

For the T(D,$\gamma$)\ce{^5He} reaction, equation (\ref{eq:emissivity}) is solved using realistic $n_D$ and $n_T$ profiles from ref.~\cite{rodriguez2022}, in which a PRD plasma is simulated using the TRANSP code~\cite{Hawryluk1980}. We then use Bosch-Hale functions to calculate the plasma reactivity $\left<v_{\mathrm{rel}} \, \sigma_\gamma\right>_{v_{\mathrm{rel}}}$ depending on the average ion temperature in each position of the plasma. The resulting poloidal map of the plasma yield for DT fusion reactions is reported in \cref{fig:DT:emissivity}, together with the rates of \gammarays that will reach the end of the NCAM collimators. The emissivity is maximum at the magnetic axis and could be best measured by a detector placed on one of the central channels, 9 or 11. Channel 10 is not considered as it hosts a magnetic proton recoil diagnostic for neutron spectroscopy~\cite{mackie2024, mackie2026, dallarosa2024} which is not compatible with a \gammaray spectrometer. Such a detector would receive $\approx8\times10^5$ $\gamma$/s if a collimator with $3$ cm diameter were to be used, and $1\times10^4$ $\gamma$/s with a $1$ cm collimator, consistent with etendue scaling like $\approx \mathrm{d}^4$.

\subsection{\ce{^{4}He}\ce{^{10}B} reaction}\label{sec:emissivity:alpha10b}

Boron is introduced in the plasma via boronization of the SPARC first wall. Detailed simulations of impurity levels and penetration inside the plasma core are still ongoing. In this work, we assume that B is present in the plasma at a $1$\% concentration, with natural isotopic abundance (\ce{^{11}B} at 80.1\% and \ce{^{10}B} at 19.9\%), and with a radial profile similar to the electron profile $n_e$ (see~\cref{fig:transp:alpha10B:n}). The isotropic nuclear reactions between $\alpha$-particles and \ce{^{10}B} can be used to study the dynamics of fusion born $\alpha$-particles. 
It is a two step reaction that can emit up to $6$ different \gammarays, with energies spanning $180$ keV to $3.85$ MeV, depending on the excited state of \ce{^{13}C} right after the reaction~\cite{kiptily2018}.

For fusion related purposes, we are interested mostly in the three \gammarays with energies between $3$ and $4$ MeV; thus, the reaction formulas can be written as:
\begin{eqnarray}\label{eq:alpha10b}
    \alpha+\ce{^{10}B} &\longrightarrow& \ce{^{14}N^*} \nonumber\\ 
    \ce{^{14}N^*} &\longrightarrow& 
    \begin{cases}
        \ce{^{13}C}+\ce{p}+\gamma(3.09~\mathrm{MeV}),\\
        \ce{^{13}C}+\ce{p}+\gamma(3.69~\mathrm{MeV}),\\
        \ce{^{13}C}+\ce{p}+\gamma(3.85~\mathrm{MeV}).
    \end{cases}
\end{eqnarray}
The complete description of the possible decay paths of \ce{^{13}C^*} can be found in ref.~\cite{kiptily2018}, which has been used to calculate the branching ratio of each \gammaray emission. These branching ratios, as well as the total cross-section of the reaction, have a strong dependency on the energy of the incident $\alpha$-particle. 
In the present work, we use the total cross-section reported in ref.~\cite{bonner1956} and the relative intensities of the three different \gammarays measured in ref.~\cite{gallmann1969}. These data cover few angles of emission (40-50 deg in~\cite{bonner1956}, 90 deg in~\cite{gallmann1969}) and a limited energy range (2.0-5.5 MeV in~\cite{bonner1956}, 1.0-3.5 MeV in~\cite{gallmann1969}). We interpolate refs.~\cite{bonner1956} and~\cite{gallmann1969} to get a cross-section for each individual \gammaray emitted in \ce{^{10}B}($\alpha$, p $\gamma$)\ce{^{13}C}. These cross-sections cover the energy range 1.0-3.5 MeV and are shown in~\cref{fig:transp:alpha10B:sigma}. The actual design of a \gammaray diagnostic for SPARC would need more detailed measurements of the \ce{^{10}B}($\alpha$,p $\gamma$)\ce{^{13}C} cross section, covering more angles and energies down to 10s of keV.

\textbf{
\begin{figure}[H] 
\centering
\subfigure[]{\label{fig:transp:alpha10B:n}
\includegraphics[width=0.9\linewidth]{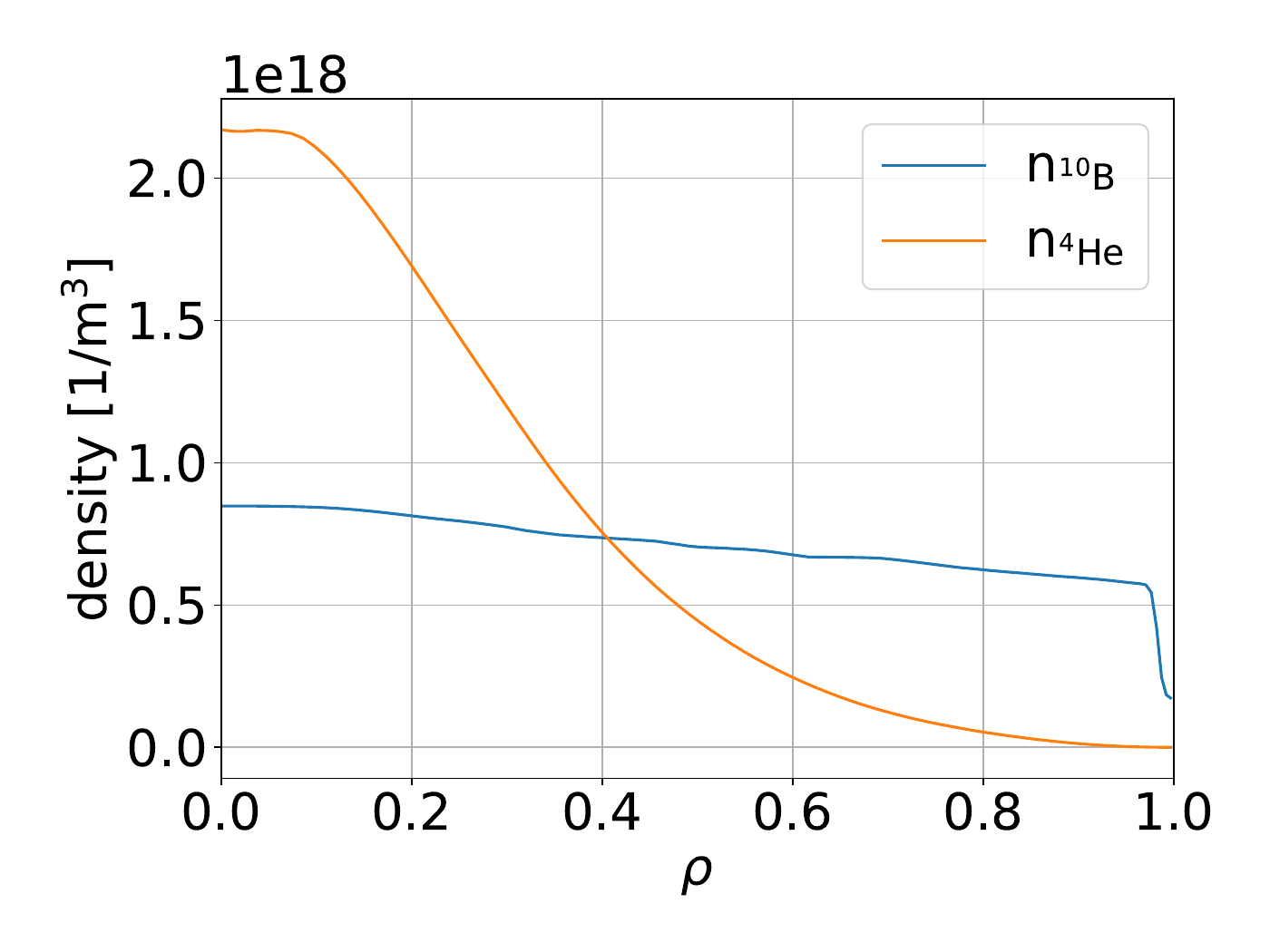}
}\\
\subfigure[]{\label{fig:transp:alpha10B:fAlpha}
\includegraphics[width=0.9\linewidth]{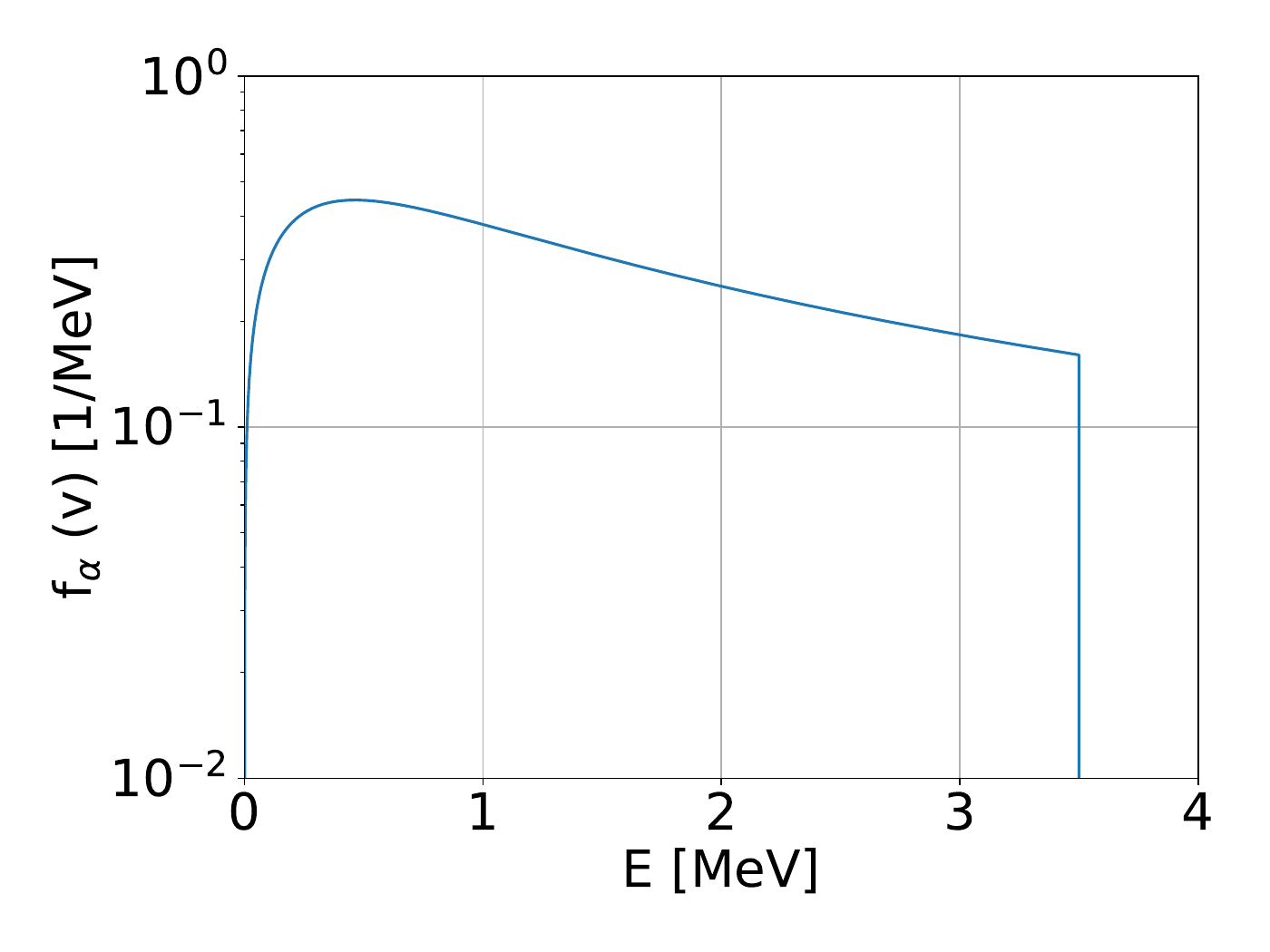}
}\\
\subfigure[]{\label{fig:transp:alpha10B:sigma}
\includegraphics[width=0.9\linewidth]{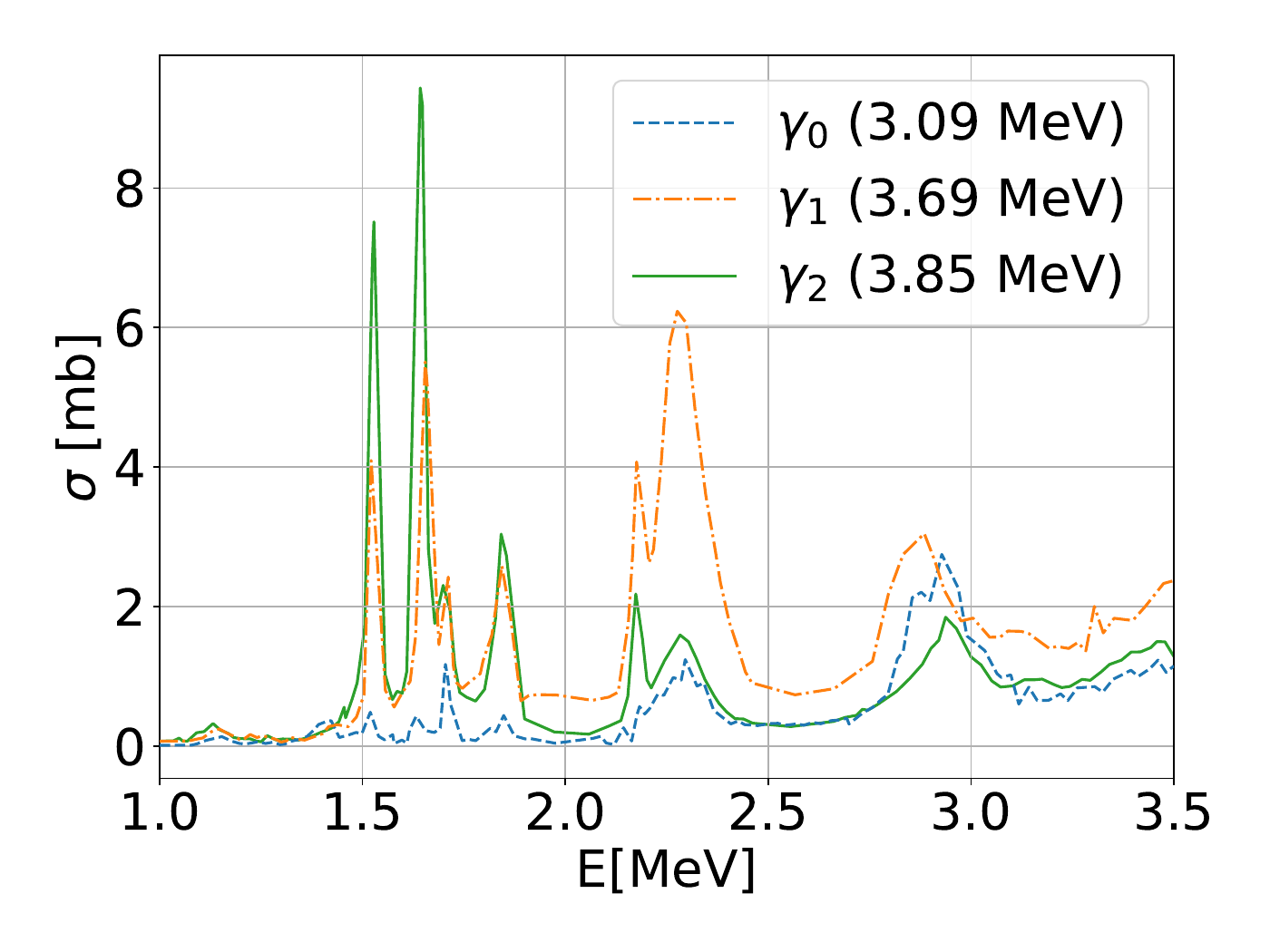}
}\\
\caption{\subref{fig:transp:alpha10B:n} From the TRANSP simulations in ref.~\cite{rodriguez2022}: density profiles of $\alpha$-particles and \ce{^{10}B} impurity (assuming B is present in the plasma at $1$\% of the electron density \ce{n_e}). \subref{fig:transp:alpha10B:fAlpha} energy distribution of $\alpha$-particles according to eq. (112) of ref.~\cite{moseev2019}. \subref{fig:transp:alpha10B:sigma} interpolation of the \ce{^{10}B}($\alpha$, p $\gamma$)\ce{^{13}C} cross-sections from refs.~\cite{bonner1956, gallmann1969}.}
\label{fig:transp:alpha10B}
\end{figure}}

\begin{figure*}[!tp] 
\centering
\subfigure[]{\label{fig:alpha:b10:emissivity:3:0:poloidal}
\includegraphics[scale=0.35]{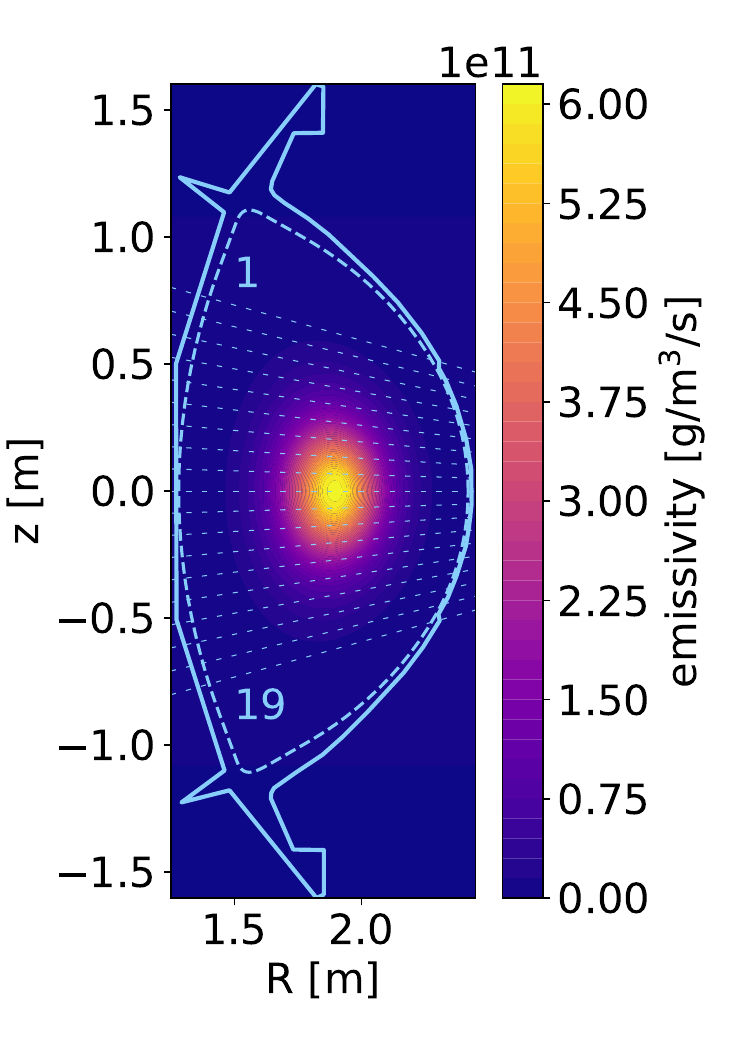}
}\subfigure[]{\label{fig:alpha:b10:emissivity:3:0:ncam}
\includegraphics[scale=0.33]{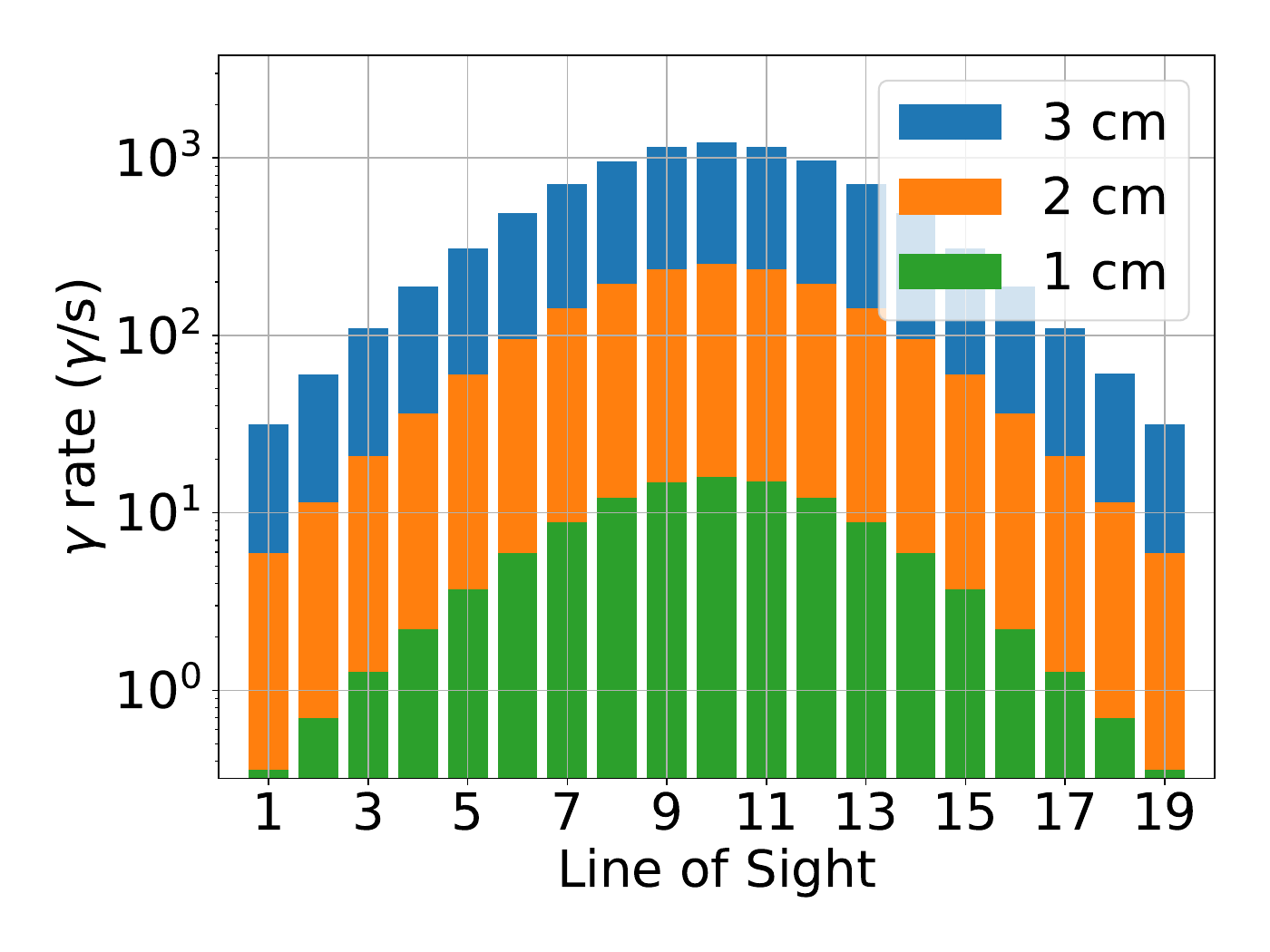}
}\\
\subfigure[]{\label{fig:alpha:b10:emissivity:3:7:poloidal}
\includegraphics[scale=0.35]{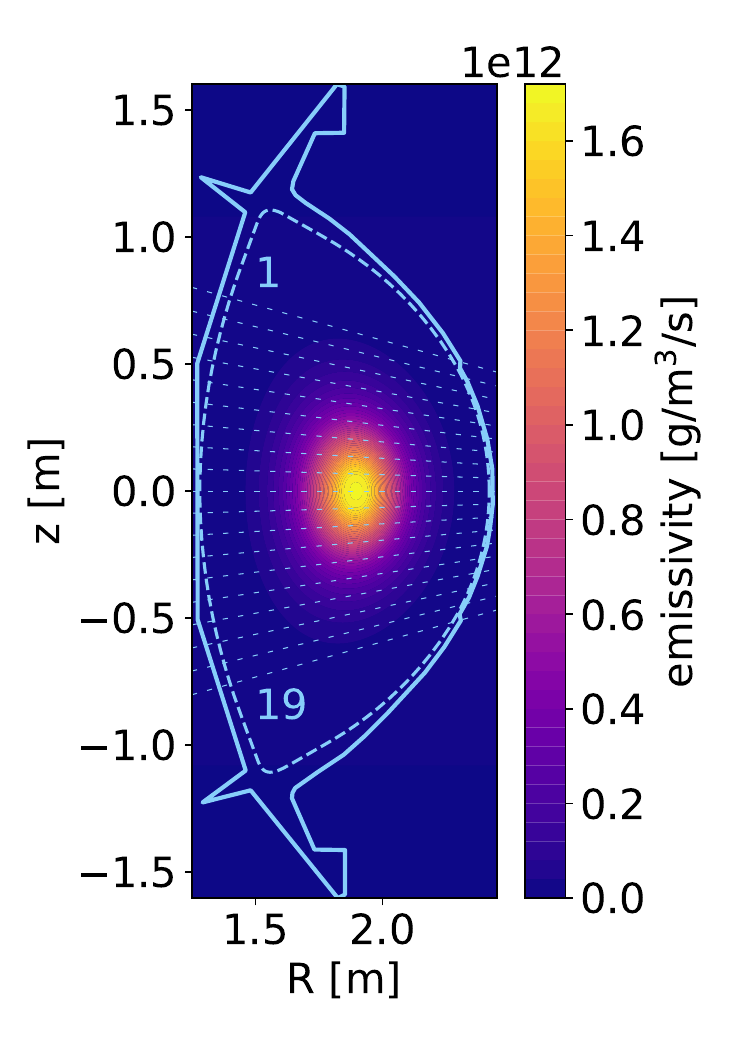}
}\subfigure[]{\label{fig:alpha:b10:emissivity:3:7:ncam}
\includegraphics[scale=0.33]{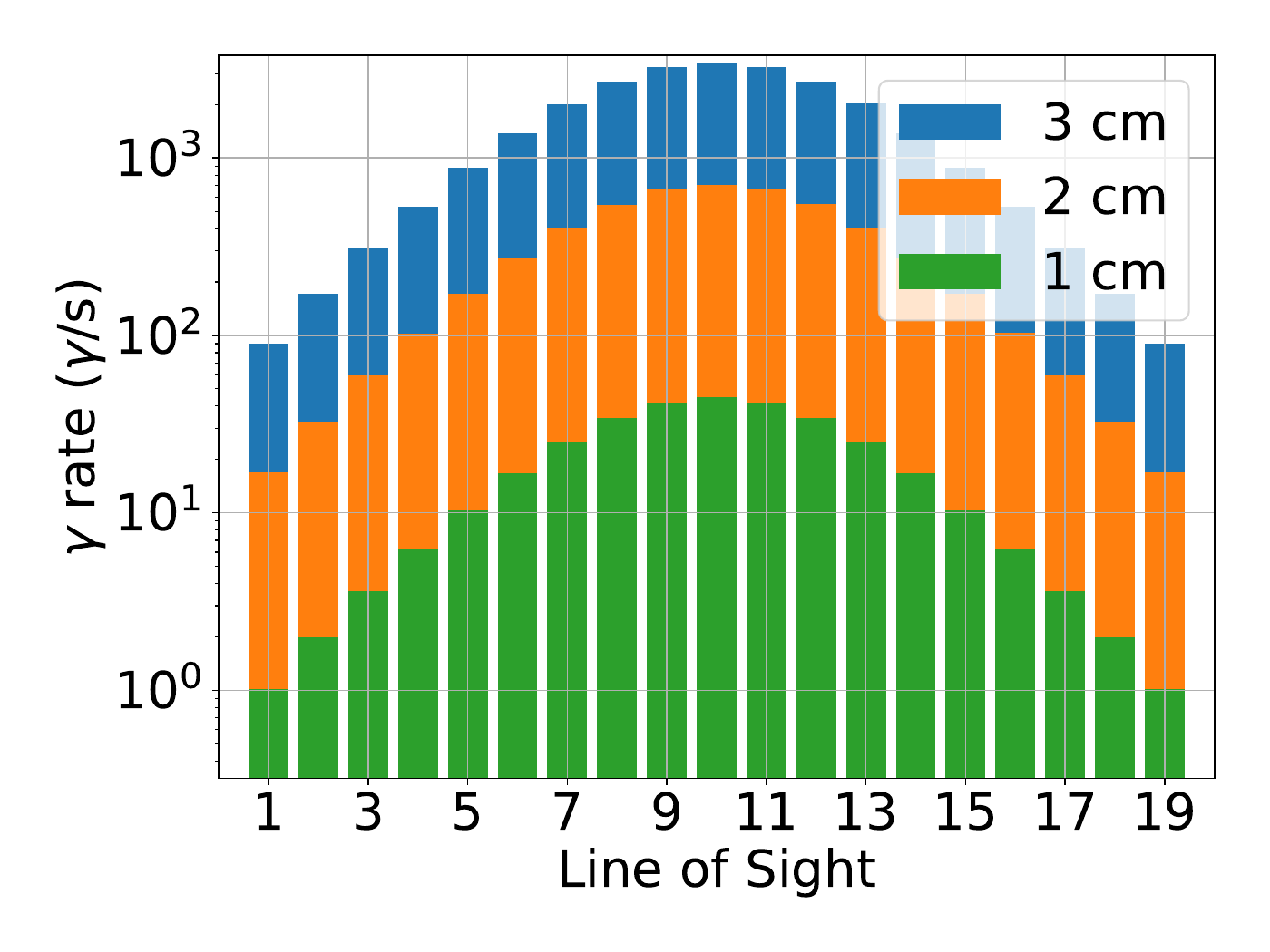}
}\\
\subfigure[]{\label{fig:alpha:b10:emissivity:3:9:poloidal}
\includegraphics[scale=0.35]{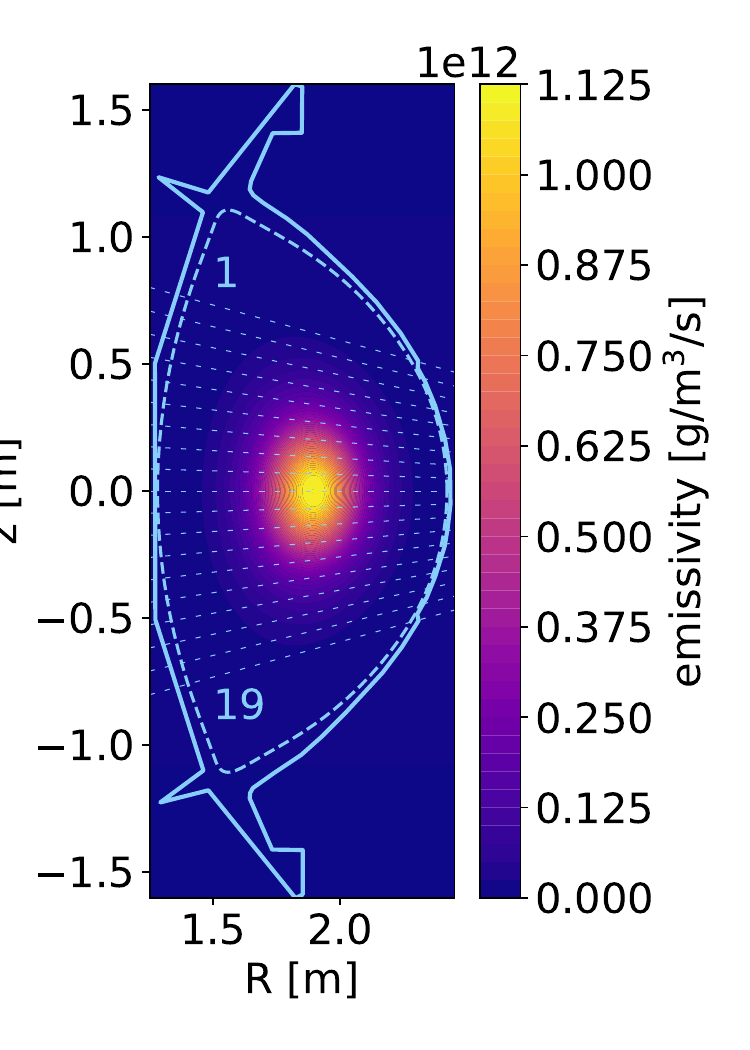}
}
\subfigure[]{\label{fig:alpha:b10:emissivity:3:9:ncam}
\includegraphics[scale=0.33]{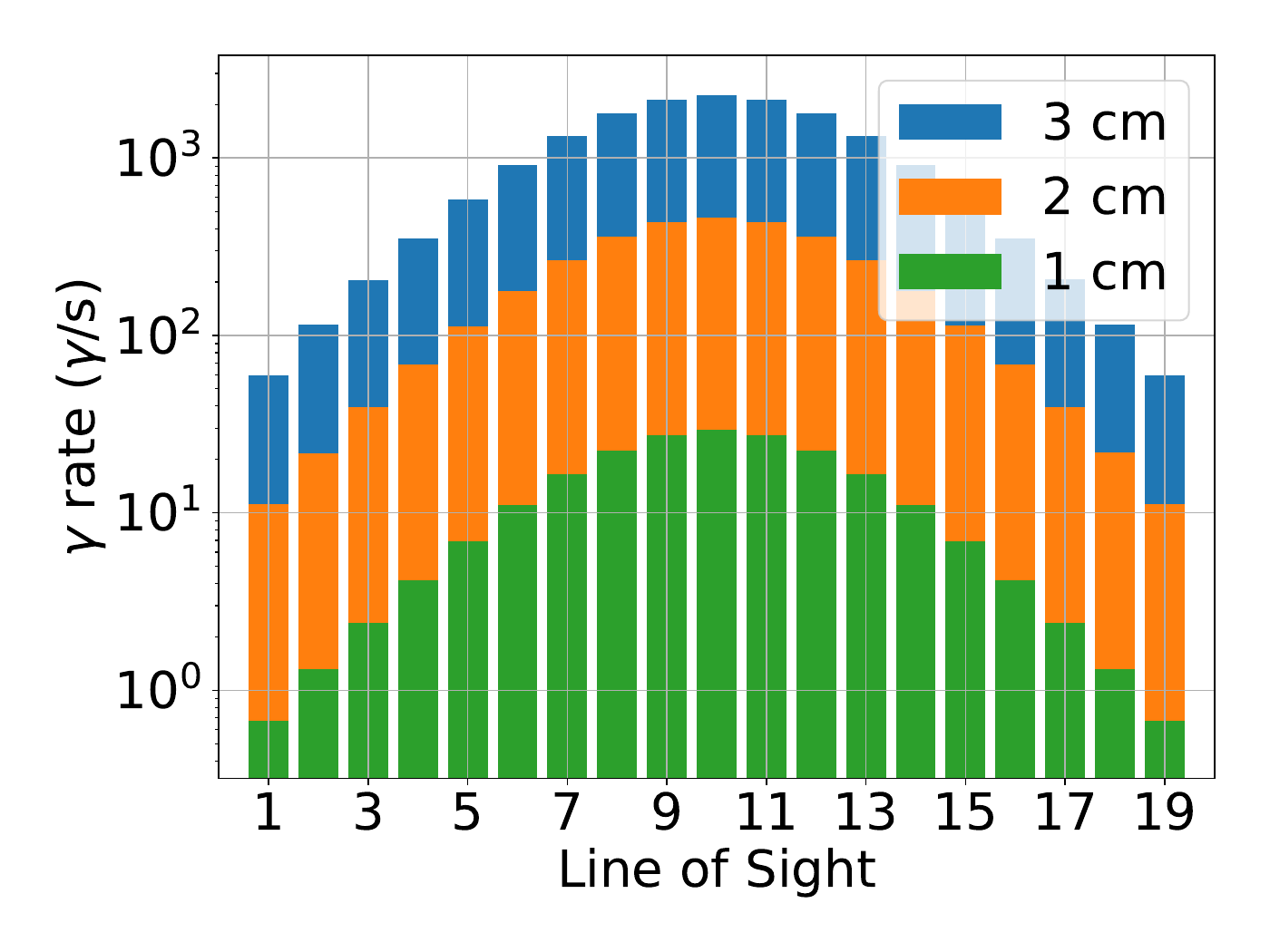}
}
\caption{Left column: poloidal profile of plasma yield for the: \subref{fig:alpha:b10:emissivity:3:0:poloidal} $3.09$ MeV, \subref{fig:alpha:b10:emissivity:3:7:poloidal} $3.68$ MeV, and \subref{fig:alpha:b10:emissivity:3:9:poloidal} $3.85$ MeV \gammaray emissions of the \ce{$\alpha$+^{10}B} nuclear reaction. First wall, LCFS and NCAM LOS are shown with light-blue with top and bottom LOS labeled. Right column: expected rate of \ce{$\alpha$^{10}B} born, \subref{fig:alpha:b10:emissivity:3:0:ncam} $3.09$ MeV, \subref{fig:alpha:b10:emissivity:3:7:ncam} $3.68$ MeV, and \subref{fig:alpha:b10:emissivity:3:9:ncam} $3.85$ MeV \gammarays at the end of the NCAM collimators as calculated with ToFu. Three collimator diameters are considered: $1$ cm (green), $2$ cm (orange), $3$ cm (blue).}
\label{fig:alpha:b10:emissivity}
\end{figure*}

\begin{figure*}[!t] 
\centering
\subfigure[]{\label{fig:d3he:toric:profiles}
\includegraphics[width=0.45\linewidth]{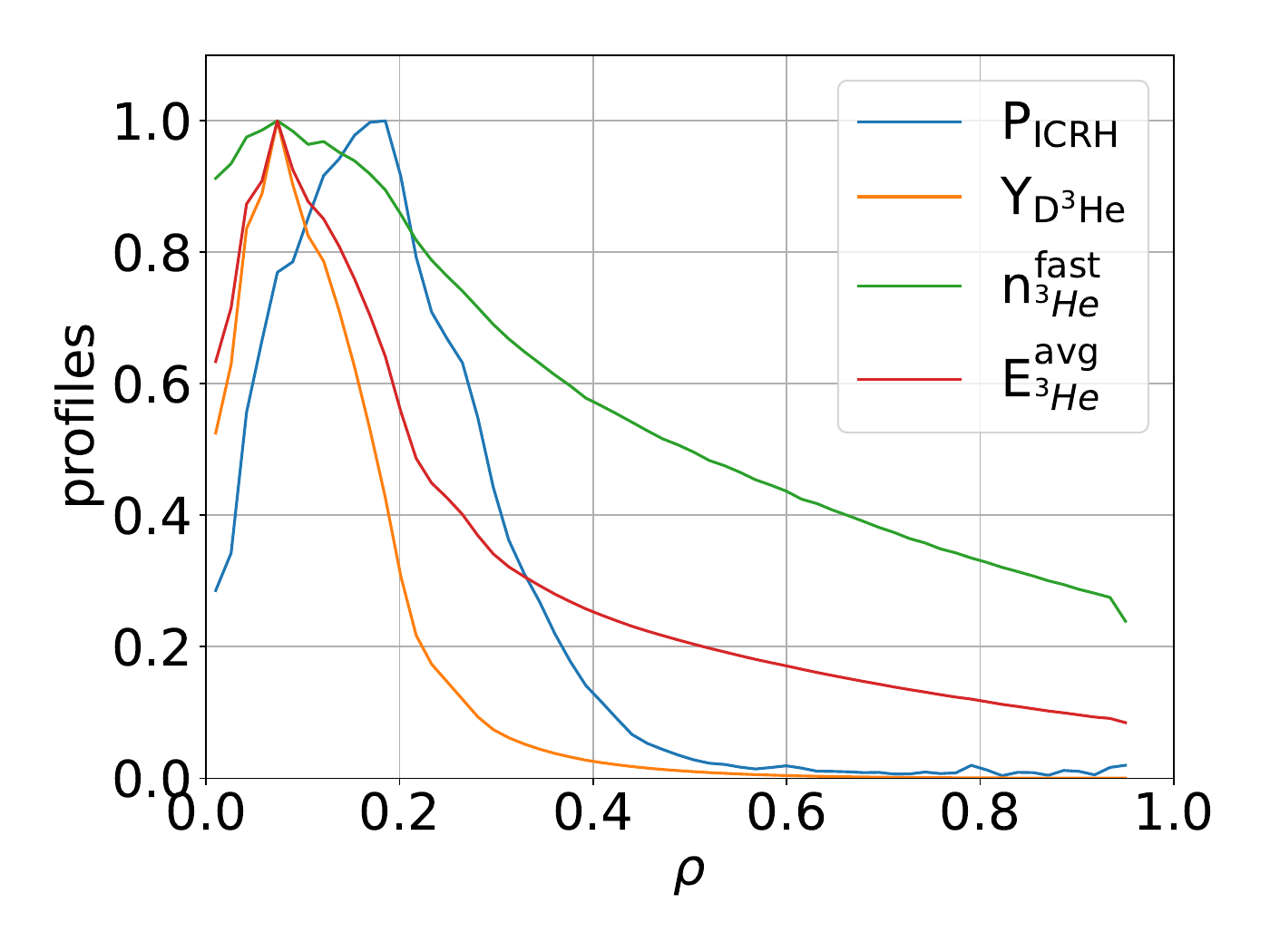}
}
\subfigure[]{\label{fig:d3he:toric:f:he3}
\includegraphics[width=0.45\linewidth]{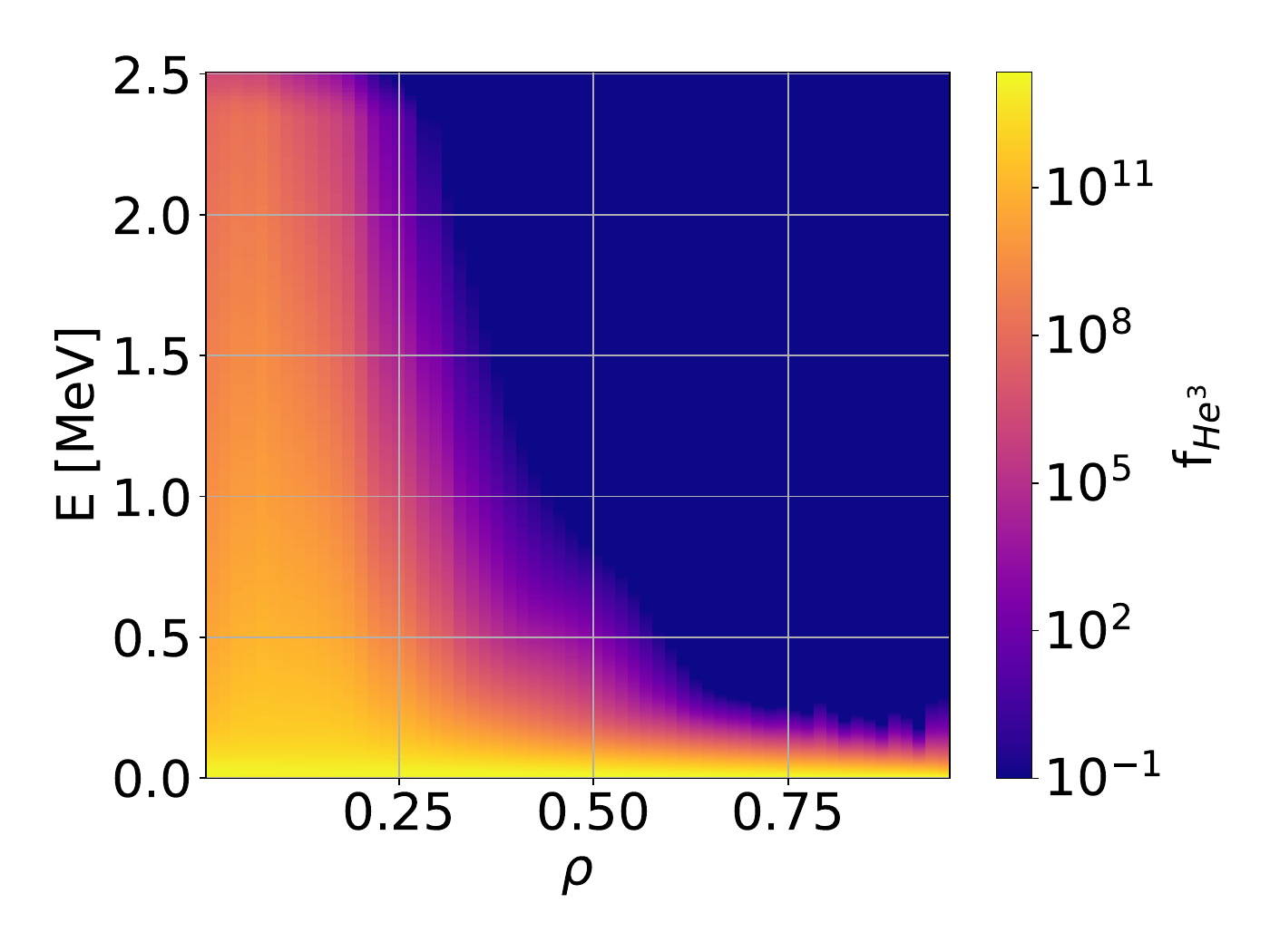}
}\\
\caption{CQL3D+TORIC simulations from ref.~\cite{mackie2026}. \subref{fig:d3he:toric:profiles} ICRH power density deposited in the plasma (\ce{P_{ICRH}}), \gammaray yield for the \ce{D^3He} reaction ($\mathrm{Y}_{\ce{D^{3}He}}$), density of the fast population of \ce{^3He} ($\mathrm{n}^{\mathrm{fast}}_{^3He}$), and average energy of the fast population of \ce{^3He} ($\mathrm{E}^{\mathrm{avg}}_{^3He}$). The maximum of all profiles is normalize to 1.\subref{fig:d3he:toric:profiles} Energy distribution of the high energy tail of \ce{^3He} at different radial locations.}
\label{fig:d3he:toric}
\end{figure*}

\begin{figure*}[!b] 
\centering
\subfigure[]{\label{fig:d3he:emissivity:poloidal}
\includegraphics[width=0.33\linewidth]{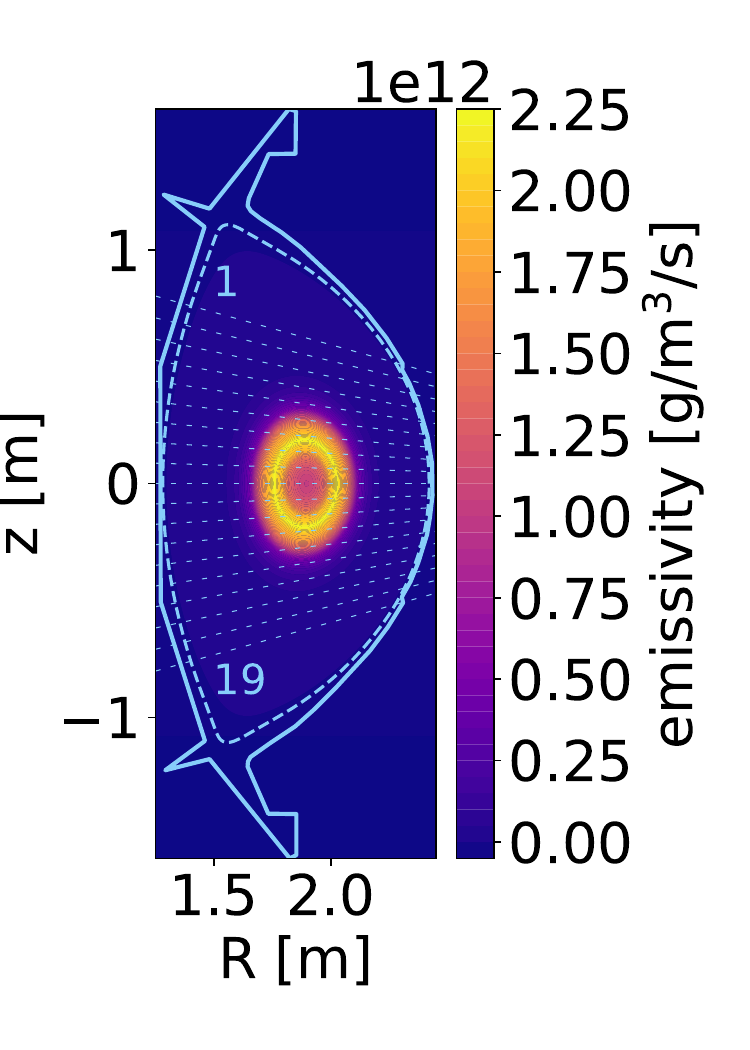}
}
\subfigure[]{\label{fig:d3he:emissivity:ncam}
\includegraphics[width=0.58\linewidth]{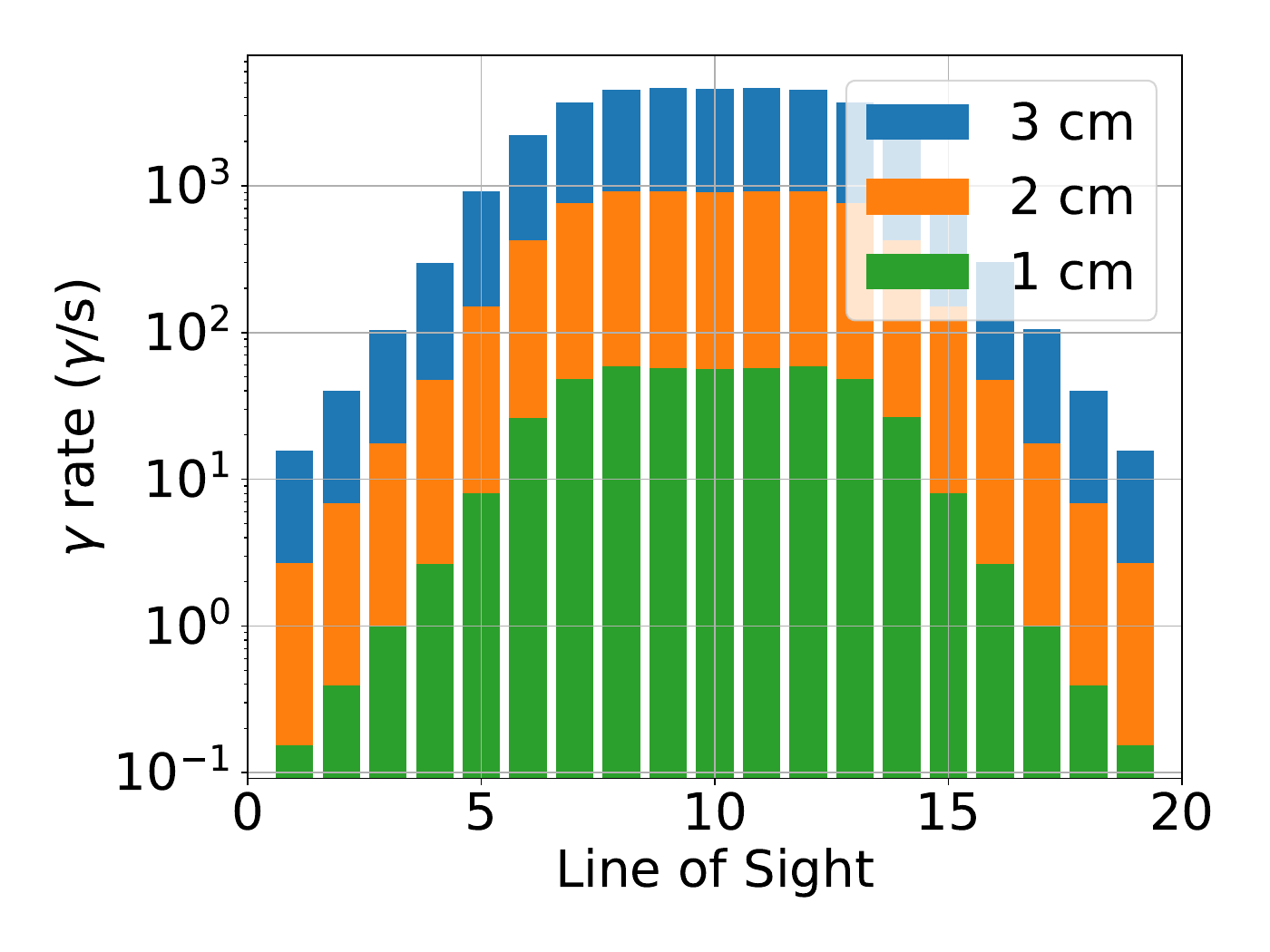}
}\\
\caption{\subref{fig:d3he:emissivity:poloidal} Poloidal profile of plasma yield for the D+\ce{^3He} fusion reaction. First wall, LCFS and NCAM LOS are shown with light-blue with top and bottom LOS labeled. \subref{fig:d3he:emissivity:ncam} Expected rate of DT born, 16.4 MeV \gammarays at the end of the NCAM collimators as calculated with ToFu. Three collimator diameters are considered: $1$ cm (green), $2$ cm (orange), $3$ cm (blue).}
\label{fig:d3he:emissivity}
\end{figure*}

Since fusion born $\alpha$-particles are emitted with an energy of $3.5$ MeV, we can approximate equation (\ref{eq:alpha10b}) as a reaction between fast ions ($\alpha$) and a fixed target (\ce{^{10}B}). Thus equation (\ref{eq:emissivity}) can be simplified as:
\begin{equation}\label{eq:emissivity:alpha10b}
\mathrm{d}\, Y_\gamma = n_{\ce{^{10}B}} \, n_{\alpha} \int_{v_{\alpha}} f_{\alpha}(v_{\alpha}) \, v_{\alpha} \, \sigma_\gamma(v_{\alpha}) \, \mathrm{d} \! v_{\alpha}.
\end{equation}
where $f_{\alpha}(v_{\alpha})$ has been set to be the slowing down distribution derived in equation (97) of ref.~\cite{moseev2019} (see~\cref{fig:transp:alpha10B}). Based on these inputs, the emissivity of the three \gammarays and their rates behind the NCAM collimators are presented in \cref{fig:alpha:b10:emissivity}. The three signals have comparable rates, with the $3.69$ MeV \gammaray being the most intense. Even assuming a perfect detector with 100~\% efficiency and no background, the expected count rates for a $1$ cm collimator would be around $45$ $\gamma$/s in the central channel of the NCAM. Integrated over a $10$ s flat-top of the PRD-like plasma and considering Poisson statistics, this would give a statistical uncertainty of $4.7$\% at best. Wider collimator diameters would increase the statistic of such a measurement and allow to study the time evolution of the $\alpha$ population during a single discharge. With $2$ and $3$ cm collimators, we can expect about $8\times10^2$ and $3\times10^3$ $\gamma$/s. Over a 10 s flat-top, these rates correspond to a Poisson uncertainty of at least $1$ and $0.5$\% over a single discharge. The actual net statistics of the $\alpha$+\ce{^{10}B} signal will depend on the detector efficiency and signal-to-background ratio, which will be covered in \cref{sec:bkgrd,sec:signal}.

\subsection{D\ce{^{3}He} fusion reaction}\label{sec:emissivity:d3he}
The most common branch of the D+\ce{^{3}He} fusion reaction produces only heavy charged particles that are confined inside the plasma. Similar to the DT fusion reaction discussed in \cref{sec:emissivity:dt}, an uncommon branch of the D+\ce{^{3}He} fusion reaction can emit a \gammaray with probability $\approx (4.5\pm1.2)\times10^{-5}$~\cite{cecil1985}. The complete scheme of this reaction is:
\begin{equation}\label{eq:d3he}
    \ce{D}+\ce{^{3}He}
    \begin{cases}
       \overset{\approx 1}{\longrightarrow} &\alpha(3.7~\mathrm{MeV})+\ce{p}(16.6~\mathrm{MeV}),\\
       \overset{\approx 4.5\times10^{-5}}{\longrightarrow} &
            \begin{cases}
                \ce{^{5}Li}+\gamma_0(16.7~\mathrm{MeV}),\\
                \ce{^{5}Li}+\gamma_1(14~\mathrm{MeV}).             
            \end{cases}
    \end{cases}
\end{equation}
Also in this case, the \gammarays are emitted with a broad energy distribution, which has been studied in ref.~\cite{cecil1985, buss1968} using the R-matrix method. 


On SPARC, ICRH will deposit its energy on the \ce{^{3}He} minority. 
This process has been studied in ref.~\cite{lin2020, mackie2026} using TORIC and CQL3D to calculate both the ICRH energy deposition and the subsequent interactions of the fast \ce{^{3}He} population with the main thermal ion species. 
In particular, the simulations performed for ref.~\cite{mackie2026} return both the profile of fast \ce{^3He} and the total \gammaray emission due to D(\ce{^{3}He}, $\gamma$)\ce{^5Li}, which are shown in \cref{fig:d3he:toric,,fig:d3he:emissivity}. 
ICRH energy is deposited off-axis, resulting in a hollow \gammaray emissivity poloidal profile. 
\Cref{fig:d3he:toric:profiles} shows that the radial profiles of the \gammaray emissivity, the density of fast \ce{^3He} accellerated by the ICRH, and the average energy of the fast \ce{^3He} all peak around $\rho\approx 0.1$.
Reconstructing the radial location of the maximum \gammaray emissivity, thus, can give direct information on where the high-energy tail of the \ce{^3He} population is located inside the plasma.
On the other hand, TORIC+CQL3D predict that the profile of the ICRH power deposition is peaked at $\rho\approx0.2$. 
\gammaray spectroscopy, then, could help validating these simulations by comparing the synthetic \gammaray emissivity, calculated using the workflow presented in this paper, with actual experimental data.

\begin{figure*}[!b] 
\centering
\subfigure[]{\label{fig:dd:emissivity:poloidal}
\includegraphics[scale=0.4]{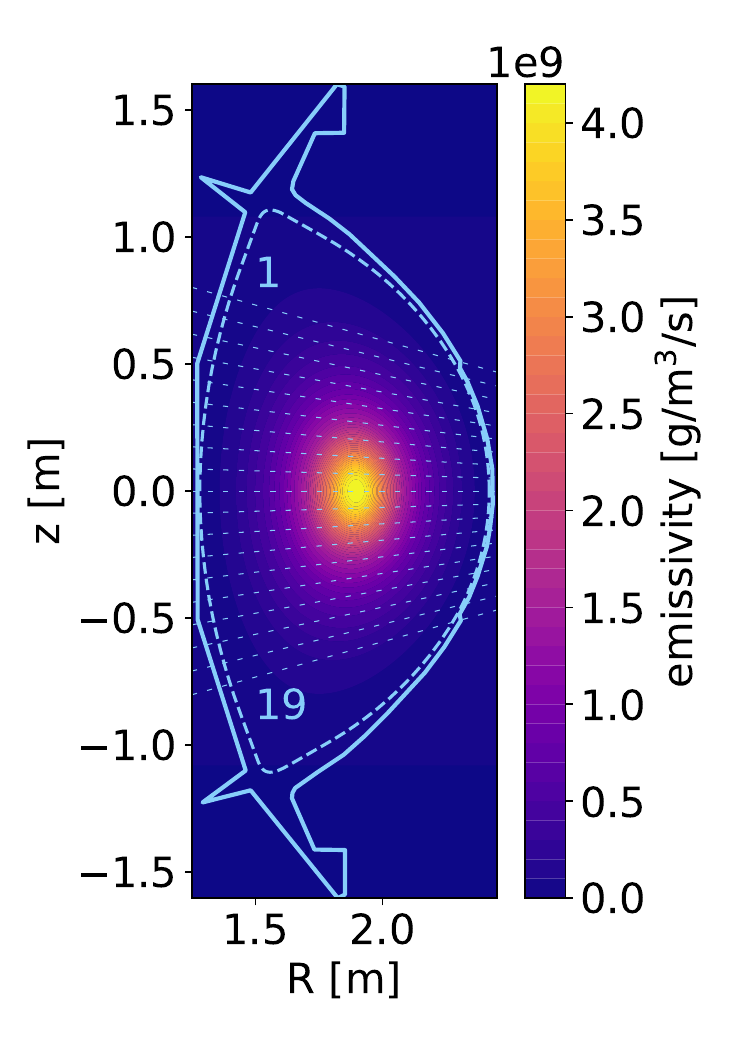}
}
\subfigure[]{\label{fig:dd:emissivity:ncam}
\includegraphics[scale=0.4]{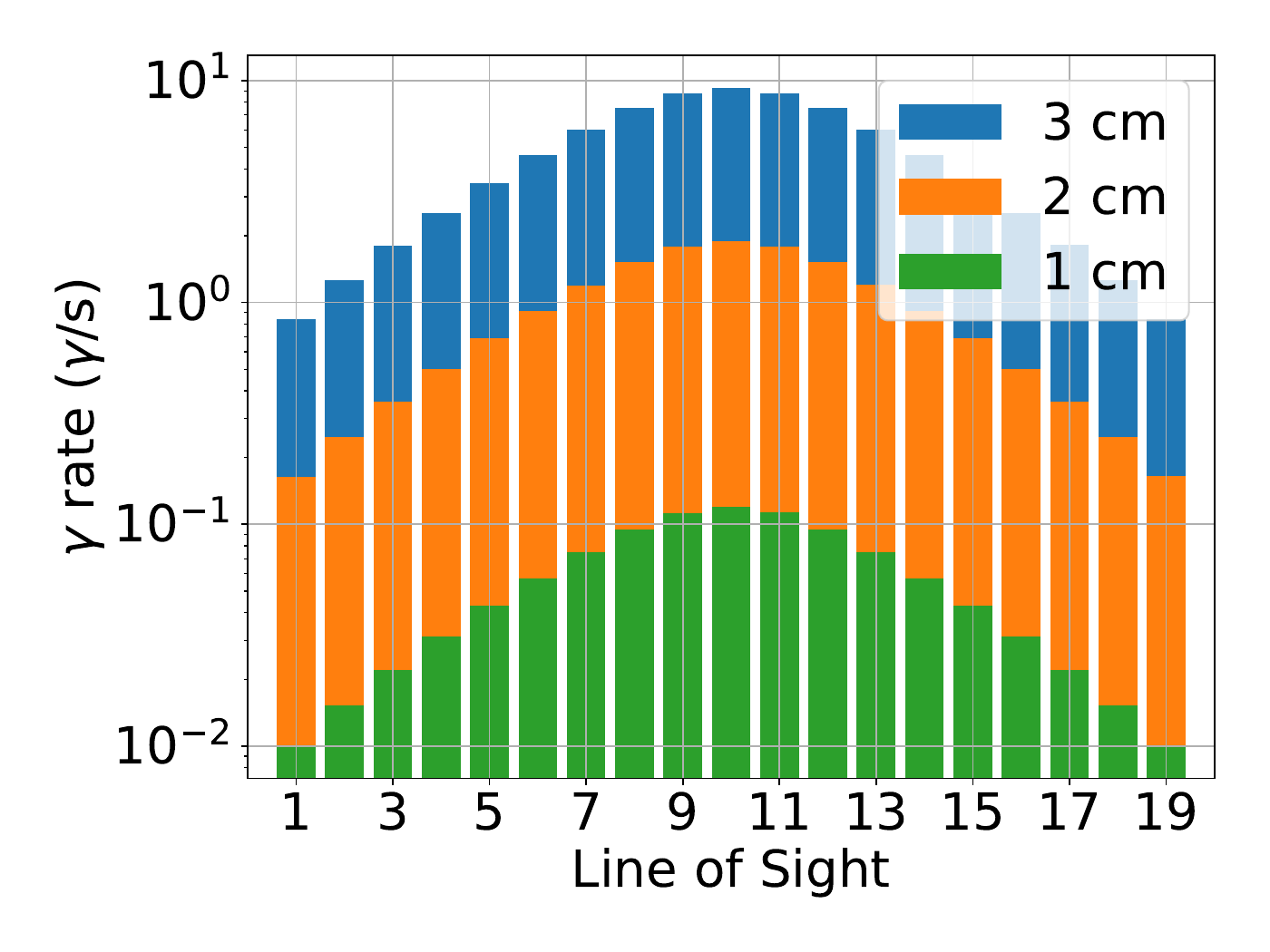}
}\\
\caption{\subref{fig:dd:emissivity:poloidal} Poloidal profile of plasma yield for the DD fusion reaction (all branches included). First wall, LCFS and NCAM LOS are shown with light-blue with top and bottom LOS labeled. \subref{fig:dd:emissivity:ncam} Expected rate of DT born, 23.8 MeV \gammarays at the end of the NCAM collimators as calculated with ToFu. Three collimator diameters are considered: $1$ cm (green), $2$ cm (orange), $3$ cm (blue).}
\label{fig:dd:emissivity}
\end{figure*}

The profile of the \gammaray reaching the end of the NCAM collimators is expected to be flat for the central LOS (7 to 13,~\cref{fig:d3he:emissivity:ncam}). 
It differs significantly from the profile of DT \gammaray, which is peaked in the central LOS (10,~\cref{fig:DT:emissivity:ncam}). 
The D(\ce{^{3}He}, $\gamma$)\ce{^5Li} to D(T, $\gamma$)\ce{^5He} ratio is below $0.8$\% in all channels and it is about $0.6$\% on channels 9 and 11, which are the best channels for measuring the DT \gammaray. 
To perform a measurement of $\pfus$ during DT operations in SPARC using \gammaray spectroscopy, then, the background due to the $\gamma$ emission of the D\ce{^{3}He} reaction can be ignored at first order. 
It is worth noting that the subtraction of the D\ce{^{3}He}  background is not trivial. 
Given the similar energies of the \gammarays emitted in DT and D\ce{^{3}He} reactions, we cannot spectrally distinguish the two reactions. 
This has two implications: first, \gammaray spectroscopy of D\ce{^{3}He} reactions will not be possible in DT plasmas. 
Second, the D\ce{^{3}He} \gammaray background must be modeled from plasma profiles, using an accurate knowledge of its spectral shape. 
A validation of this modeling effort could be performed during DD operations at a magnetic field of $12$ T, where $\gamma$-spectroscopy of D\ce{^{3}He} gammas would be favored by the fact that neutron emission will come prevalently from DD reactions and a small fraction from tritium burn up. 
During these experiments, the D\ce{^{3}He} \gammarays could also be used to diagnose ICRH power deposition.
Measurements of D\ce{^{3}He} \gammarays during DD operations have previously proved to be possible on JET~\cite{nocente2020, panontin2021, fugazza2026}.

The comparison of~\cref{fig:d3he:emissivity:ncam} and~\cref{fig:DT:emissivity:ncam} shows that different emissivity profiles (e.g. peaked in the plasma core or hollow) would have different signatures in the measurement of a $\gamma$-camera installed behind the NCAM.
The width of the D\ce{^{3}He} profile, in particular, can be used to validate CQL-3D+TORIC simulations radiofrequency heating, giving experimental information on where the radio frequency power has been deposited in the plasma.
If a full $\gamma$-camera is not available, \cref{fig:d3he:emissivity:ncam} also stresses the importance of installing a spectrometer on channels 7 or 13 for D\ce{^{3}He} \gammaray spectroscopy.
In such channels, the ratio of the D(\ce{^{3}He}, $\gamma$)\ce{^5Li} signal over the neutron background coming from DD reactions is expected to be maximum.

\subsection{DD fusion reaction}\label{sec:emissivity:dd}
The D+D has three possible branches. Two, more frequent, in which a neutron or a proton are emitted, and a third one, very exotic, in which a $23.8$ MeV \gammaray is generated, with a branching ratio of $(1.1\pm0.3)\times10^{-7}$~\cite{cecil1985:2}. The complete scheme of this reaction, then, reads:
\begin{equation}\label{eq:dd}
    \ce{D+D}
    \begin{cases}
        \overset{\approx 0.5}{\longrightarrow} &\ce{^3He}(0.82~\mathrm{ MeV})+\ce{n}(2.45~\mathrm{ MeV}),\\
        \overset{\approx 0.5}{\longrightarrow} &\ce{T}(1.1~\mathrm{ MeV})+\ce{p}(3.02~\mathrm{ MeV}),\\
        \overset{\approx 1.1\times10^{-7}}{\longrightarrow} &\ce{^{4}He}+\gamma_0(23.8~\mathrm{ MeV}).
    \end{cases}
\end{equation}
The  $\gamma$ emission from DD reactions was proposed as a diagnosis tool for D to T ratio in ref.~\cite{cecil1985:2}. 
 
We estimate plasma emissivity for this reaction using D profiles simulated with TRANSP and Bosch-Hale reactivity formulas, as detailed in \cref{sec:emissivity:dt}. The results are included in \cref{fig:dd:emissivity}. The expected \gammaray rates at the detector position are more than two orders of magnitude lower than those for the $\alpha$\ce{^{10}B} and D\ce{^{3}He} reactions, and almost 5 order of magnitudes lower than the DT \gammarays. This means that the background due to DD \gammarays can be neglected when measuring the other three \gammaray signals. Moreover, the DD \gammaray can be spectrally distinguished from the DT and D\ce{^3He} \gammarays. The DD contribution can eventually be subtracted from the other \gammarays by fitting the DD spectrum measured above 17.5 MeV, as done in ref.~\cite{dalmolin2024, rebai2024, marcer2025}. 
\section{Neutron background}\label{sec:bkgrd}
Scintillation materials such as \labr are sensitive to neutrons and in this section we present the expected neutron-induced background during DT operations in SPARC. Most of the neutron signal will come from $14.1$ MeV neutrons born in DT fusion reactions. On top of that $2.5$ MeV neutrons will also be emitted in DD reactions, which represent a second order correction to our calculations and will not be included in the present work. An extensive study of DD and DT neutron interactions with \ce{LaBr_3} has been conducted in Ref.~\cite{cazzaniga2013, cazzaniga2015}. The first source of background comes from neutrons that reach the detector and interact directly with the scintillation crystal. The second contribution comes from prompt $\gamma$-rays emitted in (n,$\gamma$) reactions with materials in the torus hall or in the collimation structure. Finally, the third contribution would come from neutron activation of the crystal and the surrounding materials. With some scintillation materials it is possible to distinguish between counts generated by neutrons or by \gammaray~\cite{ball2024}. Such technique would allow to exclude the first source of neutron-induced background from the postprocessing of the measured data. However, at present, we are not aware of any successful application of pulse shape discrimination (PSD) to distinguish \gammaray from neutron interactions with \labr crystals. For the scope of this synthetic study, then, we will consider the neutron-induced background to be the sum of all three components here described.

In \cref{sec:bkgrd:fluxes}, we will present the expected background levels for a \ce{LaBr_3} detector installed behind the NCAM collimators without neutron attenuation. In~\cref{sec:bkgrd:attenuator} we present neutron attenuation strategies to enable a \ce{LaBr_3} detector to perform $\gamma$-ray spectroscopy even in the high neutron fluxes produced by SPARC.

\subsection{Neutron-induced fluxes at detector position}\label{sec:bkgrd:fluxes}
\begin{figure}[H] 
\centering
\includegraphics[scale=0.34]{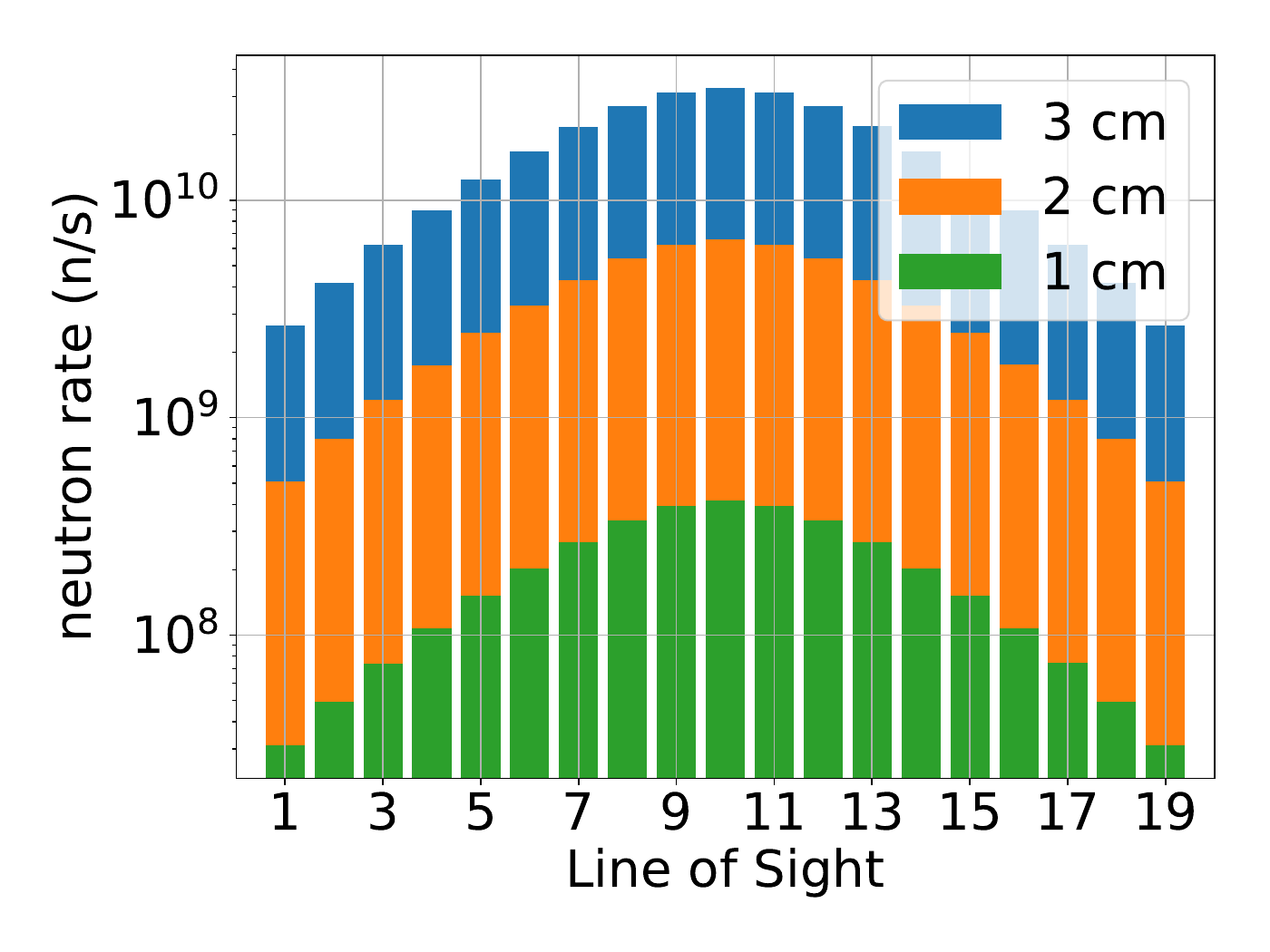}
\caption{Expected rate of DT-born, unscattered, 14.1 MeV neutrons at the end of the NCAM collimators as calculated with ToFu. Three collimator diameters are considered: $1$ cm (green), $2$ cm (orange), $3$ cm (blue).}
\label{fig:DT:n:emissivity:ncam}
\end{figure}

\Cref{fig:DT:n:emissivity:ncam} shows the rates of $14.1$ MeV reaching the end of the NCAM collimators during a PRD as calculated by ToFu. The profile is peaked at the central LOS of the NCAM, which has a collimator diameter of 3 cm and is expected to receive $3.3\times10^{10}$ n/s. The workflow used to perform these calculations is the same used in \cref{sec:emissivity:dt}. The neutron rates reaching the end of the central NCAM collimator for two different plasma scenarios, PRD and Q>1, and two collimator configurations, $D=1$ and $3$ cm, are reported in \cref{tab:n:rates}. The minimum neutron rate is $3.0\times10^7$ n/s, expected for Q>1 scenario with $D=1$ cm collimator. The results of this workflow have been validated with a high fidelity Monte Carlo simulation of SPARC in ref.~\cite{wang2025}. Experience on past machines, such as JET~\cite{panontin2021, nocente2020, nocente2021, dalmolin2024}, showed that gamma spectroscopy with \ce{LaBr_3} is possible only if the total count rate on the detector is at most around $5\times10^5$ Cps. That means that such high neutron rates reaching the detector alone would saturate an inorganic scintillator, and thus require appropriate attenuation.

The contribution of (n,$\gamma$) reactions has been estimated using a full Monte Carlo approach. 
In this section, we present the prompt-gamma emission generated by neutrons that interact with materials in the tokamak hall and the diagnostic laboratories wall, which will affect any $\gamma$-ray measurement regardless of the detector technology and the neutron attenuator deployed. 
Then, in \cref{sec:bkgrd:attenuator}, we cover the prompt-gamma emission from the neutron attenuator material. 
For the contribution coming from the torus hall, we used the OpenMC model presented in Ref.~\cite{wang2024, wang2025}, which implements a 60 deg model of the SPARC torus hall and diagnostic laboratories wall with high-fidelity. 
The complexity of the model reduces the statistics of the tally, thus limiting us to use an 8 energy group. 
This detailed study of the (n,$\gamma$) signal has been conducted only for the midplane LOS, which represents the worst case scenario for all channels of the NCAM. 
The total rates reaching the detector are summarized in \cref{tab:n:rates}, showing that direct neutron fluxes are $10$ times higher than prompt-gamma fluxes coming from SPARC. 

\begin{Table}
\centering
\captionof{table}{\label{tab:n:rates} Rate of neutron and prompt-gammas born in the torus hall, which will reach the end of the NCAM collimators.}
\smallskip
\begin{tabular}{l l r r}
\toprule
& collimator & \ce{R_n} [n/s] & \ce{R_{(n,$\gamma$)}} [$\gamma$/s]\\
\midrule
{\it PRD}&$D=3$ cm & $3.3\times10^{10}$ & $1.7\times10^9$\\
         &$D=1$ cm & $4.2\times10^8$ & $4.1\times10^7$\\
{\it Q>1}&$D=3$ cm & $2.3\times10^9$ & $1.2\times10^8$\\
         &$D=1$ cm & $3.0\times10^7$ & $2.9\times10^6$\\
\bottomrule
\end{tabular}
\end{Table}

\begin{figure}[H]
\centering
\includegraphics[scale=\subfigsize]{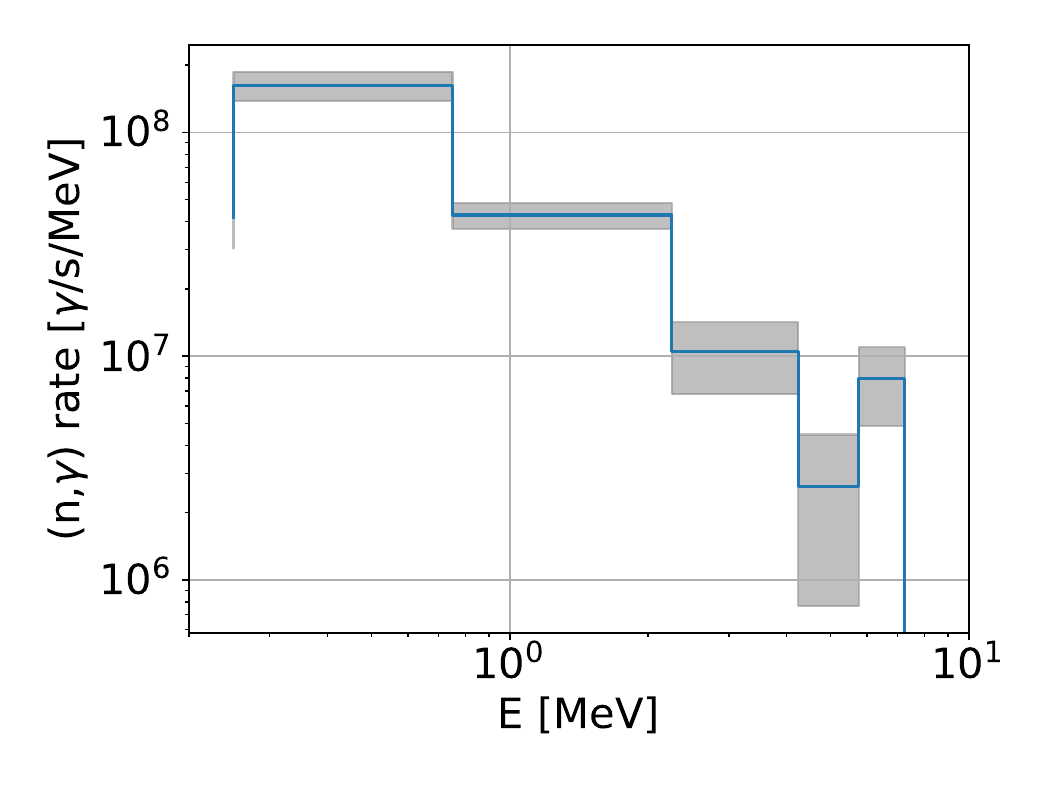}
\captionof{figure}{OpenMC simulations of the energy spectrum of prompt $\gamma$-rays generated in the torus hall that reaches the end of the NCAM collimators during a PRD. Shaded region shows interval of confidence.}
\label{fig:ng:flux}
\end{figure}

The spectrum of the (n,$\gamma$) reaching the detector is shown in \cref{fig:ng:flux}. The simulations show prompt-gammas with energies as high as $8$ MeV, but are unable to capture any event between $8$ and $20$ MeV due to lack of statistics. 
Neutron-induced background in this energy region has been measured on JET~\cite{dalmolin2024, rebai2024} during DT operations.
Such a signal is also visible in the simulations of the prompt-gammas generated by a neutron attenuator presented in~\cref{sec:bkgrd:attenuator}. 
However the magnitude of this high energy emission is quite low and can be subtracted from the $\gamma$ signal with a fit on the experimental data. Recently, the high energy (n,$\gamma$) background has been investigated using deterministic neutronics simulation with weight windows~\cite{colombi2026:sofe, colombi2026:jinst}, showing promising implications for background subtraction application in \gammaray spectroscopy.
In the energy region below $8$ MeV, the background due to prompt-gammas would dominate any measurements, hiding all $\gamma$ emission coming from the plasma. 

Due to the high neutron emissivity in the compact SPARC volume, the prompt-gamma background from the torus would be in excess of $5\times10^5$ $\gamma$/s even during a Q>1 discharge with a $D=1$ cm collimator.
To reduce the expected radiation flux reaching the detector, in~\cref{sec:bkgrd:attenuator} we scope an attenuator for $\gamma$ studies.
The background resulting from neutrons directly reaching the detector can be moderated using a neutron attenuator that favors the transmission of $\gamma$-rays over neutrons, thus improving the gamma-to-neutron ratio after the attenuator. 
On the other hand, neutrons interacting with the attenuator will undergo (n,$\gamma$) reactions, thus decreasing the gamma-to-prompt-gamma ratio after the attenuator. Moreover, there is no possibility to distinguish the plasma \gammarays from the neutron-induced prompt-gamma, not even if we were to use a scintillation material that has PSD capabilities. 
Then, signal extraction must rely on the analysis of the energy spectrum and requires an accurate background subtraction, which severely limits the possibility of measuring certain $\gamma$ reactions. 

\begin{Table}
\centering
\captionof{table}{\label{tab:alpha:b10:signal:noise} Signal to noise ratio at the end of the HDPE neutron attenuator for $\gamma$-rays born in \ce{$\alpha$^{10}B} nuclear reactions. Background is assumed to be composed only by (n,$\gamma$) coming from the torus hall. Integration time is set to 10 s.}
\smallskip
\begin{tabular}{l r r }
\toprule
E$_\gamma$ [MeV] & D = 1 cm & D = 3 cm \\
\midrule
3.09 & $0.09 \pm 0.02$ &  $1.1 \pm 0.2$ \\
3.68 & $0.53 \pm 0.19$ &  $6.4 \pm 2.3$ \\
3.85 & $0.35 \pm 0.13$ &  $4.2 \pm 1.5$ \\
\bottomrule
\end{tabular}
\end{Table}

Considering mono-energetic $\gamma$-rays, such as $\alpha$+\ce{^{10}B} emission, their peak can be extracted from the spectrum only if the signal-to-noise ratio is high enough, say at least $3$. 
For instance, the signal-to-noise ratio of the $\alpha$+\ce{^{10}B} reaction is reported in \cref{tab:alpha:b10:signal:noise}.
The calculation compares the $\gamma$ signal from~\cref{sec:emissivity:alpha10b} with the prompt-gamma background coming from the torus hall, which is the minimum background we would experience in such a measurement even if we were to develop a detector insensitive to direct neutron interactions. 
The best energy resolution of \labr crystals at $662$ keV is usually considered to be 2.8 \%, which corresponds to a resolution of $1.4$\% at $3.5$ MeV and a FWHM of about $40$ keV.
From~\cref{fig:ng:flux}, we expect about $N_{(n,\gamma)} \approx 2.96\times10^7$ prompt-gammas coming from the torus hall in an energy interval of $40$ keV centered around $3.5$ MeV during $10$ s.
We assume the noise on the prompt-gamma background to be equal to the Poisson uncertainty (about $\sqrt{N_{(n,\gamma)}} \approx 5.44\times10^3$). 
\Cref{tab:alpha:b10:signal:noise} shows that, right after the collimator, it could be in principle possible to distinguish the $\alpha$\ce{^{10}B} signal from the prompt-gamma background if the B concentration is at least $1$\% in the core, the collimator diameter is $D=3$ cm and we integrate over the full $10$ s PRD plasma. 
This motivates follow-up studies on the prompt-gamma background to increase the energy resolution of its spectrum, and on the transport of \ce{^{10}B} impurities to the plasma core in a high B-field machine such as SPARC.

\subsection{Neutron attenuator scoping}\label{sec:bkgrd:attenuator}
The high fluxes expected for the neutron-induced background at the end of the NCAM collimators during a PRD would prevent any kind of $\gamma$ spectroscopy with inorganic scintillators.  
In this section we investigate if a neutron attenuator installed behind the NCAM collimators could decrease the particle rates hitting the detector to about $5\times10^5$ particles/s, to avoid large pile-up in the detector and enable spectroscopy. 
Here a particle can be a neutron, a prompt-gamma or a $\gamma$-ray with energy above the low energy threshold of the digitization chain, which we found to be $0.1$ MeV for \ce{LaBr_3}~\cite{panontin2024}. 
We consider an attenuator slab made of high density polyethylene (HDPE) with a density of $1.08$ \ce{g/cm^3} and an atomic composition for: C (33\%) and H (66\%).
The wall between the torus and the diagnostic laboratories is $2.5$ m thick, and hosts two $60$ cm thick Al cylinders that define the aperture of the field of view of the NCAM~\cite{wang2025}. 
We are interested in an attenuator thickness below $1.3$ m, so that it could be integrated inside the NCAM collimators in future design iterations without perturbing the field of view of the detector.

\begin{figure}[H]
\centering
\includegraphics[scale=\subfigsize]{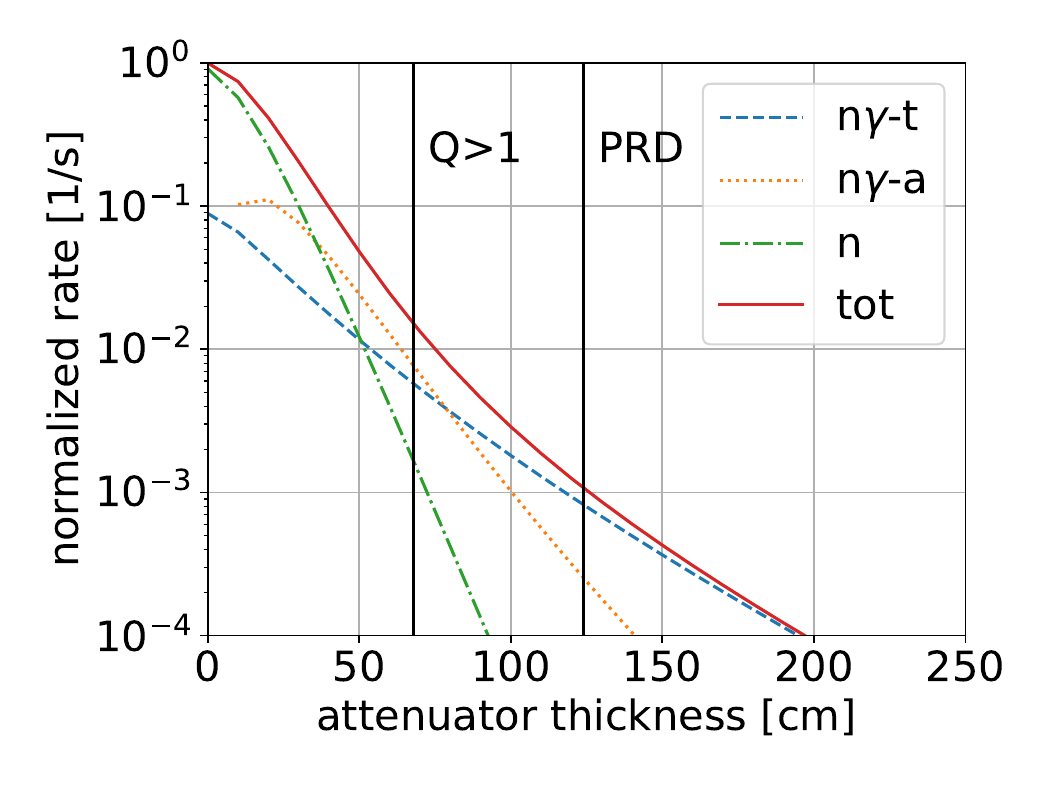}
\captionof{figure}{
Attenuation of the neutron-induced background by a 250 cm HDPE slab.
Direct neutrons (n), prompt-gammas emitted in the torus hall (n$\gamma$-t) and by the attenuator (n$\gamma$-a) are shown, together with their sum (tot). 
All rates are normalized to the total initial rate. 
For a collimator diameter of 1 cm, the thickness necessary to attenuate the total rate to $5\times10^5$ particle/s is shown for both PRD and Q>1 scenarios.}
\label{fig:attenuator:length:scan}
\end{figure}

The attenuator has been modeled as a large ($50$ cm width x $50$ cm height x $250$ cm depth) slab of material installed behind the NCAM collimators. 
A dedicated OpenMC model has been implemented to study the evolution of the radiation fluxes inside the collimator, with tallies defined every $10$ cm.
Each tally counts the current of particles flowing through the surface in the direction that goes from the torus to the detector, which corresponds to a surface tally of the particle flux.
The model takes as sources the neutron and $\gamma$-ray fluxes emerging from the back aperture of the NCAM collimator, as calculated in \cref{sec:emissivity,sec:bkgrd:fluxes}. 
This approach allows assessment of the flux coming directly from the torus; however, it does not capture the cross-talk between different NCAM channels, nor any shine-through contributions coming from the wall or the collimator inserts.

\begin{Table}
\centering
\captionof{table}{\label{tab:hdpe:tickness} Thickness of a HDPE attenuator necessary to reduce the neutron-induced background rate to $5\times10^5$ particles/s for collimator diameters (D) of 1 and 3 cm.}
\smallskip
\begin{tabular}{l r r }
\toprule
Plasma scenario & $D = 1$ cm & $D = 3$ cm \\
\midrule
PRD & $124$ cm & $246$ cm \\
Q>1 & $68$ cm & $158$ cm \\
\bottomrule
\end{tabular}
\end{Table}

\begin{figure*}[t] 
\centering
\subfigure[]{\label{fig:measured:spectrum:dtg}
\includegraphics[scale=\subfigsize]{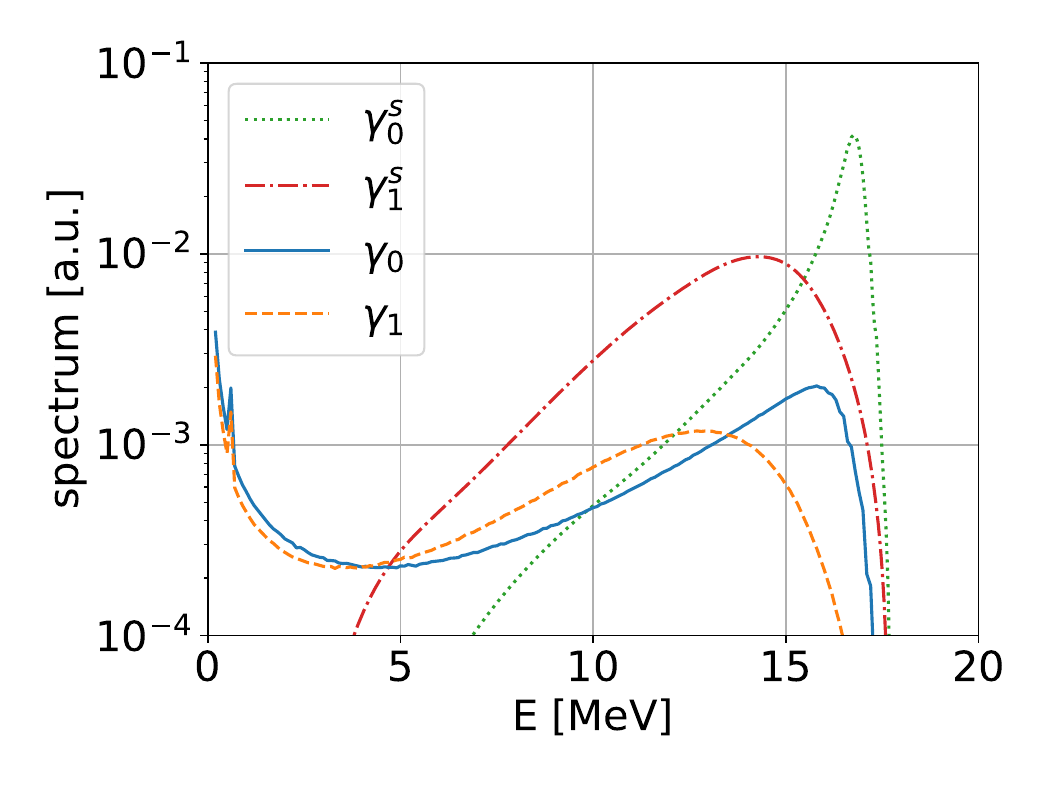}
}
\subfigure[]{\label{fig:measured:spectrum:all}
\includegraphics[scale=\subfigsize]{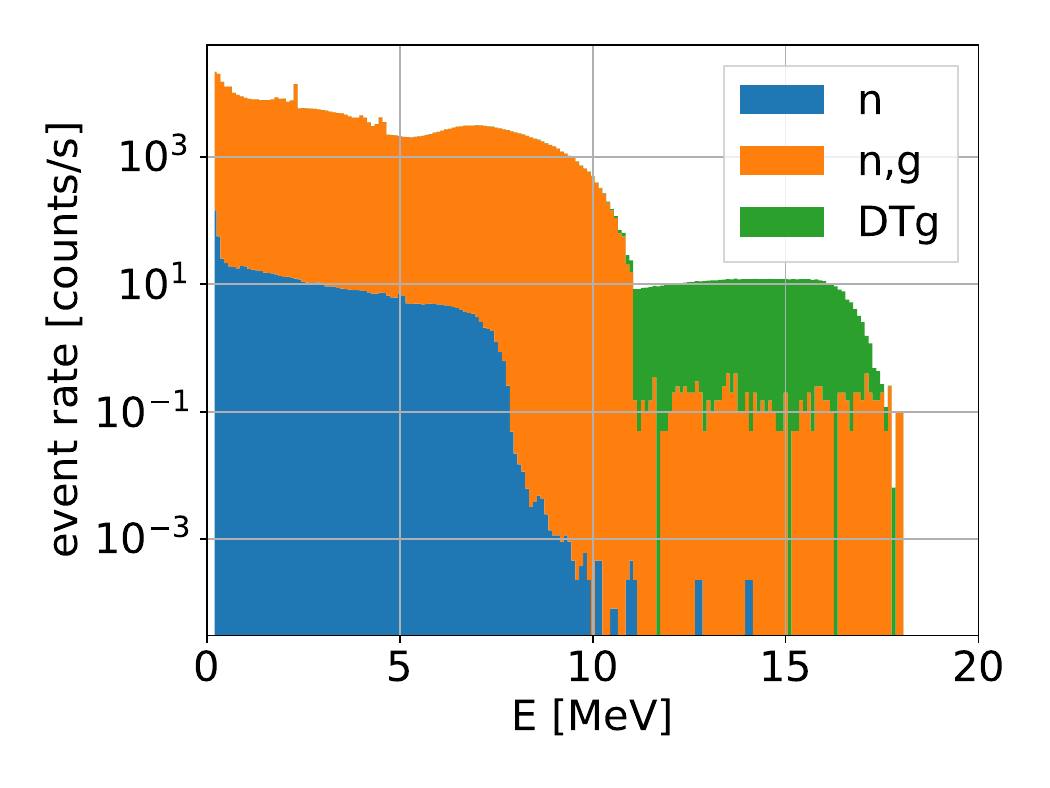}
}
\caption{\subref{fig:measured:spectrum:dtg} \ce{LaBr_3} detector response function to $\gamma_0$ and $\gamma_1$ emitted in DT fusion reactions. Both spectra are normalized per source photon emitted by the plasma. The spectrum of the DT emission in the plasma (source $\gamma_0$ and source $\gamma_1$) is the one measured in Ref.~\cite{rebai2024} and is normalized to unity. \subref{fig:measured:spectrum:all} Total spectrum measured by a \ce{LaBr_3} detector during a PRD scenario.}
\label{fig:measured:spectrum}
\end{figure*}

Two plasma scenarios have been considered: PRD and Q>1, as well as two collimator sizes: $D=1$ and $3$ cm. 
For each case, the thickness of HDPE necessary to reduce the rate hitting the detector to $5\times10^5$ particle/s is reported in \cref{tab:hdpe:tickness}. 
A collimator diameter of $D=3$ cm would require an attenuation thickness above $1.3$ m for all scenarios considered: Q>1 and PRD. 
On the other hand, if $D=1$ cm, the attenuator would be well within the 1.3 m limit in both scenarios. 
For this reason, in section~\ref{sec:signal} we consider only the $D=1$ cm case.
It is worth noting that the (n,$\gamma$) flux coming from the torus hall is expected to dominate the background reaching the detector. 
During a PRD, for example, the (n,$\gamma$) from the torus hall are predicted to be $\approx76$\% of the total background, (n,$\gamma$) from the HDPE contribute are $\approx24$\% and neutrons are less than $0.3$\%.
This shows that the prompt-gamma background would be the limiting factor of a $\gamma$-spectrometer on SPARC.
It is also worth mentioning that we considered also LiH as an attenuator material, which was proposed in ref.~\cite{fugazza2026:1} to perform \gammaray spectroscopy on SPARC using a \labr detector.
According to our OpenMC simulations, the thickness of material necessary to attenuate the signal to the desired rate is always in excess of $1.3$ m. 
For Q>1 and a collimator of $D=1$ cm, we estimate the need for $157$ cm of LiH. 
For any other scenario, the necessary thickness of LiH was estimated to be greater than the collimator length ($250$ cm).

\section{Measured signal}\label{sec:signal}
We study the performance of a \ce{LaBr_3} cylindrical detector as a $\gamma$-ray spectrometer used to reconstruct $\pfus$ on SPARC.
The dimensions of the detector should be large enough for the detector to cover the entire solid angle defined by the collimators and to stop the fast electrons ($E_{e^-}$ up to $\approx20$ MeV) traveling through the scintillation crystal as a result of $\gamma$-ray interaction with the detector. 
Given the geometrical constraints of the NCAM collimators, reported in \cref{sec:emissivity} and the stopping power of \ce{LaBr_3} crystals, from the ESTAR database~\cite{berger2005}, each size of the crystal should be of at least $2.5$ inch. 
In this work we have assumed cylindrical crystals of size $3\times6$ inch$^2$, similarly to what has been used at JET~\cite{curuia2017, nocente2020}.

The detector response to $\gamma$-rays has been computed with MCNP, while the energy deposition of neutrons have been estimated using Grasshopper~\cite{danagoulian2021}. 
Neutrons can interact with \ce{LaBr_3} via different channels, generating both electron and heavy charged particles: protons, deuterons and $\alpha$-particles. 
Each type of particle has in turn a different yield of scintillation photons. 
When reporting the energy deposition spectrum calculated via a radiation transport code, we need to convert the energy transferred to heavy particles into electron equivalent. 
In this work, we use a conversion factor of 0.796 for protons, 0.648 for deuterons and 0.353 for $\alpha$-particles, as measured in~\cite{crespi2009, cazzaniga2015, cazzaniga2016}. 

In our simulations, we consider neutrons directly reaching the detector, prompt-gammas generated both in the torus hall and in the neutron attenuators, and the DT $\gamma$ emission. 
The response to DT $\gamma$-rays is shown in \cref{fig:measured:spectrum:dtg}, compared with their source spectra. 
The total efficiency is estimated to be about 91\% for both $\gamma_0$ and $\gamma_1$. 
Observing the spectral features of the two simulations, we recognize a strong scattered component below $2$ MeV, with a $511$ keV peak coming from pair production in the collimator and attenuator materials. 
Furthermore, the measured spectra of both $\gamma_0$ and $\gamma_1$ are slightly shifted at lower energies, and the detector efficiency above $11$ MeV becomes $60$\% for $\gamma_0$ and $51$\% for $\gamma_1$. 

The expected measured spectrum is shown in \cref{fig:measured:spectrum:all}. 
The n-induced background dominates the measurement in the energy region below $11$ MeV, hiding the $\gamma$ signal.
Above the $11$ MeV threshold, the prompt-gamma background predicted by our simulations is $6$ times less intense than the DT \gammaray signal, thus gamma spectroscopy is possible using an appropriate background subtraction. 
The expected total count rate at the detector, obtained integrating over the whole spectrum above $100$ keV is $\approx4.5\times10^5$ counts/s, which is within the spectroscopic capabilities of \labr detectors. 
It is interesting to report, that considering the signal reduction due to the attenuator, the detector efficiency and the $11$ MeV low-energy cut, a \labr detector would measure only $6$\% of the DT \gammaray emerging from the collimator.

\begin{figure*}[!t] 
\centering
\subfigure[]{\label{fig:statistics:dtg:prd}
\includegraphics[scale=\subfigsize]{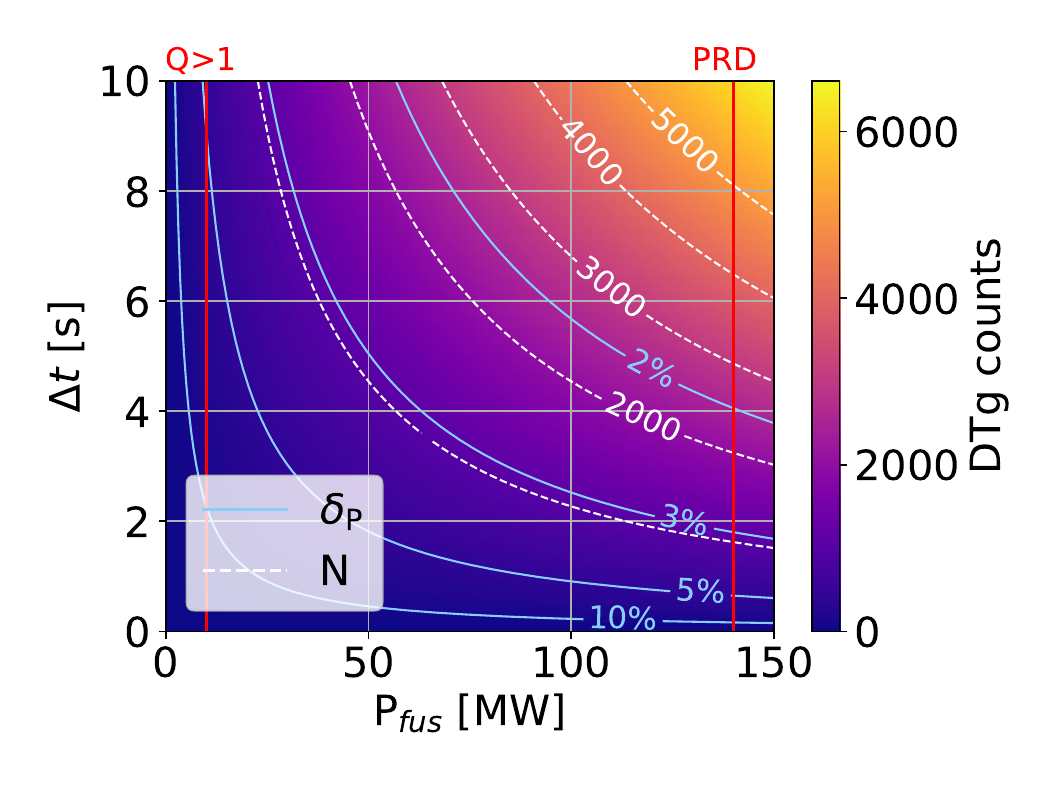}
}
\subfigure[]{\label{fig:statistics:dtg:q1}
\includegraphics[scale=\subfigsize]{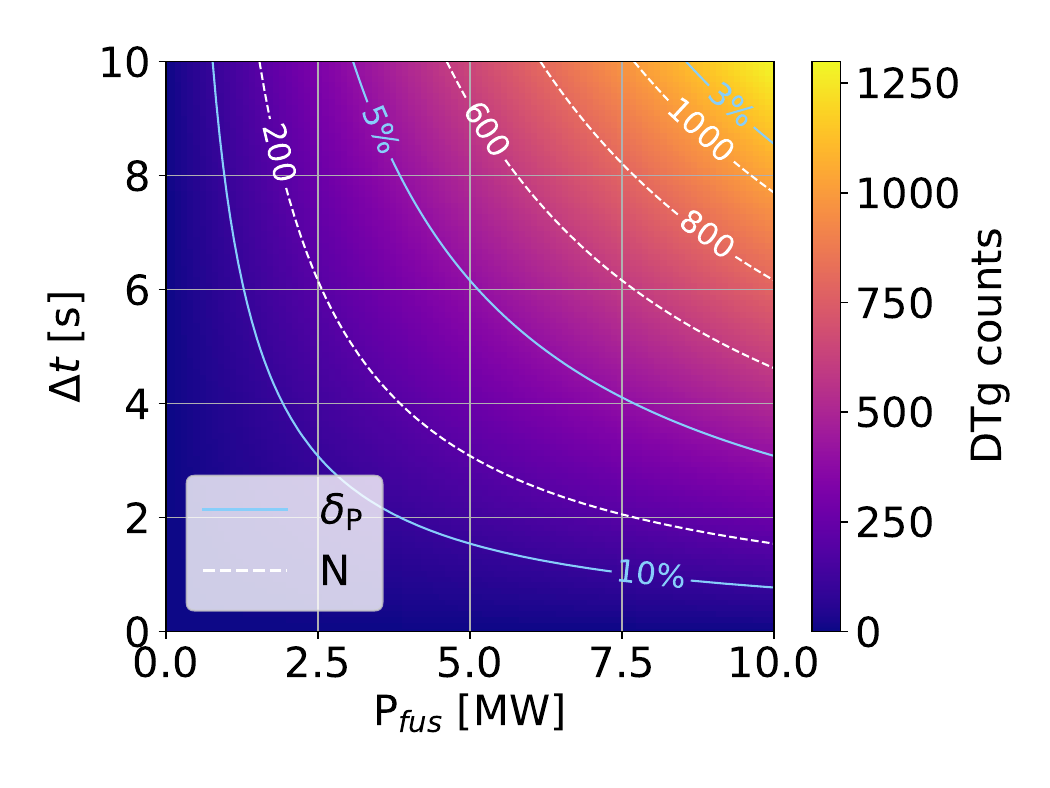}
}
\caption{Total DT $\gamma$-rays that a \ce{LaBr_3} detector is expected to measure above $10$ MeV in a time interval $\Delta t$. \subref{fig:statistics:dtg:prd} considers $124$ cm of HDPE neutron attenuator for PRD scenarios. \subref{fig:statistics:dtg:q1} considers $38$ cm of HDPE neutron attenuator for Q>1 scenarios. Contour lines for total DT $\gamma$-ray counts over the neutron background (N) and the Poisson statistics uncertainty related to the measurement ($\delta_P$) are also shown.}
\label{fig:statistics:dtg}
\end{figure*}

We scope the statistics of the DT $\gamma$-ray measurements in this energy range for different $\pfus$ and time of integration $\Delta t$ in \cref{fig:statistics:dtg}. 
Two configurations for the neutron attenuator are considered. 
In \cref{fig:statistics:dtg:prd}, 124 cm of HDPE are placed in front of the detector to optimize the measurement for a PRD scenario (see \cref{sec:bkgrd:attenuator}), where the plasma is supposed to generate up to $140$ MW of power. 
The total uncertainty on the reconstruction of $\pfus$ can be estimated as the sum in quadrature of the uncertainties coming from the Poisson statistics of the \gammaray spectroscopy, the spectral fitting, the reconstruction of the plasma profile and the cross section data. 
We want to keep the total uncertainty below $10$\%, which means that each source of uncertainty should be around $5$\%.
For what concerns the statistical uncertainty on the total number DT $\gamma$-rays counted, assuming this is governed only by Poisson statistics, the $5$\% target  corresponds to about 400 counts for the DT \gammaray signal. 
Then, for a $140$ MW plasma with $124$ cm of HDPE attenuator, $\pfus$ could be theoretically reconstructed with $0.6$ s resolution in the central LOS. 

For lower fusion power we could still reconstruct \ce{P_{fus}}. Considering the Q>1 scenario, the plasma is supposed to generate up to $10$ MW of power.
With $124$ cm of HDPE attenuator, we could still reach $5$\% statistical uncertainty over $10$ s. 
We can optimize the attenuator size and use only $64$ cm of HDPE (see \cref{sec:bkgrd:attenuator}). 
In this case, we could improve the statistical uncertainty to $\approx3$\% over $10$ s and potentially reconstruct \ce{P_{fus}} with a $10$\% total uncertainty if other sources of uncertainty are reduced below $6$\%.


According to ref.~\cite{mackie2026}, the systematic uncertainty in reconstructing $\pfus$ for SPARC due to tomographic reconstruction of the NCAM geometry can be kept within $6$\%, by using appropriate tomographic algorithms. 
Thus, the main limiting factors on the reconstruction of \ce{P_{fus}} from \gammaray spectroscopy will be the uncertainty on the cross-section for the DT fusion reaction and the fit of the DT \gammaray spectrum. 
As mentioned in~\cref{sec:emissivity:dt}, the branching ratio of the \gammaray branch has been measured with a $20$\% uncertainty in ref.~\cite{dallarosa2024}. 
Moreover, the measurements of this branching ratio conducted by multiple experiments differ by up to a factor $20$.
Performing \gammaray spectroscopy on SPARC, then, could contribute to reduce the uncertainty on these cross-sections and prepare for using this technique to measure \ce{P_{fus}} on the ARC tokamak.
On the other hand, the reconstruction (either via fit or via unfolding) of the source spectrum of DT \gammaray is an ill-posed problem, where multiple solutions can agree with the experimental data within statistical uncertainties.
\Cref{fig:DT:emissivity:ncam} shows that a \labr detector would change significantly the spectrum of the DT \gammarays, thus limiting the confidence with which we can perfor such an analysis.
The conceptual design of a {\it electron recoil} detector for magnetic confinement fusion, called MERGS~\cite{kunimune2026, fensterheim2026, huber2026}, is showing more promising results. 
According to its current design stage, MERGS might be capable to operate on SPARC without any neutron attenuator, thus having a better posed response function to high energy \gammarays and preserving more informations about the original DT \gammaray spectrum.

\begin{figure}[H] 
\centering
\includegraphics[width=\linewidth]{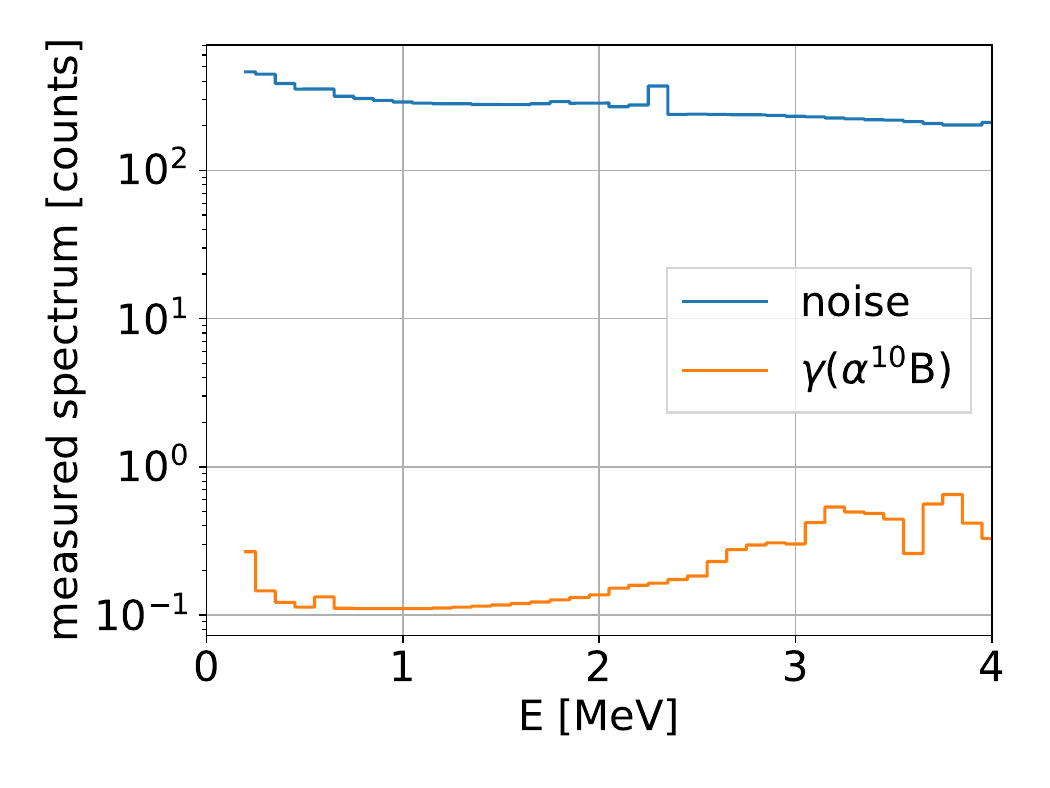}
\caption{Expected spectrum integrated for $10$ s of the $\alpha$\ce{^{10}B} \gammaray as measured by a \labr detector behind $124$ cm of HDPE attenuator ($\gamma$($\alpha$\ce{^{10}B})). Magnitude of the noise expected for the neutron-induced background in that energy range (noise).}
\label{fig:AB:measure:labr}
\end{figure}

Using the same diagnostics geometry (D$=3$ cm and $124$ cm of HDPE), we can also study how the $\alpha$\ce{^{10}B} reaction would be measured by a \labr detector. 
The energy of the \gammarays emitted are all below $11$ MeV, where the prompt-gamma background is expected to dominate the measurement. 
\Cref{fig:AB:measure:labr} shows the expected spectrum measured over $10$ s of a PRD plasma flat-top with an energy binning of $100$ keV.
The three characteristic peaks between $3$ and $4$ MeV are no longer visible and the spectrum becomes a continuum between $0.1$ and $4$ MeV.
The expected statistics is also low, with a total of $10$ counts integrated under the whole spectrum.
\Cref{fig:AB:measure:labr} also reports the magnitude of the noise expected in the same energy range, calculated as the square root of the neutron-induced background, i.e. assuming it follows a Poissonian statistics.
The $\alpha$\ce{^{10}B} signal is expected to be roughly three order of magnitude lower than the noise level.
These results cast a shadow on the capability of a \labr to extract the $\alpha$\ce{^{10}B} signal from the neutron-induced background, even integrating over the whole life of SPARC.
A possible solution can be found in \cref{tab:alpha:b10:signal:noise}: a D$=3$ cm collimator would increase the signal-to-noise by an order of magnitude.
As already pointed out, a \labr detector would not be able to work behind a large collimator aperture, due to the excessive neutron-induced background.
This result further motivates the study of alternative detection technologies, such as MERGS, capable of sustaining higher incident rates than \labr and thus operating with a D$=3$ cm collimator.

\section{Conclusions}
This paper studies the $\gamma$-ray emission during DT operations on SPARC and scopes the opportunity of measuring it with traditional \ce{LaBr_3} detectors. We present a workflow capable of estimating the \gammaray signal on SPARC starting from high fidelity plasma profiles calculated with TRANSP and realistic ICRH energy deposition calculated with CQL-3D+TORIC. The workflow solves the optical radiation transport using ToFu, and the interaction of the \gammarays with attenuation materials and detectors using Monte Carlo codes, such as MCNP and OpenMC. We identify a possible position for a $\gamma$ spectrometer behind the collimation system of the SPARC neutron camera.

DT $\gamma$-rays are expected to have intensities high enough to be measured during both Q>1 and PRD discharges, that will generate from $10$ MW to $140$ MW of power. Due to the ICRH scheme targeting \ce{^3He} minority in $B = 12$ T scenarios, the background due to \ce{D^3He} $\gamma$-rays will represent up to $0.8$\% background in the DT measurements, which can be neglected at first order. The intensity of the \ce{D^3He} emission, however, suggests there could be an opportunity for a dedicated study of the \ce{D^3He} $\gamma$-ray intensity in DD plasmas, as a diagnostics tool for energy deposition of the ICRH system. We also consider DD $\gamma$-rays, as a mean to study the D to T fuel ratio, showing that their statistic is so scarce it would require the integration over many discharges to get a significant signal.

We studied the neutron-induced background that would affect a $\gamma$-measurement performed with traditional \ce{LaBr_3} detectors. 
The OpenMC code is used to solve the neutron transport from the torus to the detector with a Monte Carlo approach. 
Direct neutron and prompt-gammas fluxes reaching the end of the neutron camera are expected to exceed the spectroscopic capabilities of a detector for collimator diameters of $1$ cm or larger in both PRD and Q>1 scenarios.
With total neutron yields during a PRD that are expected to be $10$ times higher than those experienced at JET and a plasma volume $5$ times smaller, SPARC will have a much higher neutron emission density than any previous magnetic fusion device. 
Furthermore, the neutron-induced background will mostly come from prompt-gamma generated by neutrons undergoing (n,$\gamma$) reactions in the torus.
This requires the development of new strategies for neutron attenuation that go beyond the LiH used on JET.
We scope a HDPE attenuator that could allow operations of a \ce{LaBr_3} detector installed behind the central collimator of the SPARC neutron camera during both PRD and Q>1 scenarios. 
The solution presented could be integrated inside the collimation system in future design activities.

A detailed response function of \ce{LaBr_3} detectors to $\gamma$-rays and neutrons has been calculated with MCNP.
Neutron-induced backgrounds are expected to dominate any measurement below $11$ MeV, in line with what has been experimentally detected on JET.
The spectral shape of the DT $\gamma$-rays is calculated. 
A $3\times6$-\ce{inch^2} (diameter x height) cylindrical \ce{LaBr_3} detector would have a $60$\% and $51$\% efficiency above $11$ MeV for DT $\gamma_0$ ($E_{\gamma_0} \approx 16.7$~MeV) and $\gamma_1$ ($E_{\gamma_1} \approx 13.5$~MeV) respectively.
The total efficiency of the HDPE attenuator and the \labr detector is estimated to be around 6\%.
The total counts of the DT $\gamma$-ray measurement are predicted to have a $3$\% statistical uncertainty or better over the a $10$ s flat-top of a PRD. 
In order to reconstruct $\pfus$ with a total uncertainty of about $10$\% from \gammaray spectroscopy, it is necessary to decrease the uncertainty on the branching ratio of the DT fusion reaction, which currently is of the order of $20$\% in individual measurements. 
A \gammaray diagnostics on SPARC could contribute to improving confidence in the branching ratio.
Such a measurement would be necessary and preparatory to the deployment of \gammaray diagnostics on future fusion power-plants, such as ARC, as an independent and complementary system to certify the Q performance alongside neutron spectroscopy.

Finally, we scope the statistics of the $\alpha$\ce{^{10}B} \gammaray emission, showing that a \labr detector would not be capable of distinguishing it from the neutron-induced, prompt-gamma background coming from the torus hall. Our findings, however, suggest that the signal-to-noise ratio might be high enough for a detector capable of operating with a D$=3$ cm collimator. Dedicated studies on impurity transport in high magnetic fields are necessary to confirm this result. 

\subsection*{Acknowledgements}
This work has been supported by Commonwealth Fusion Systems.

To A.D.R my star, my perfect silence.

We would also like to extend our gratitude to S. Segantin and C. Wink for the stimulating conversations on radiation transport simulations.

\printbibliography

\end{multicols}
\end{document}